\documentclass[preprint,preprintnumbers,amsmath,amssymb,floatfix, aip, pop]{revtex4-1}

\usepackage[dvips, pdftex]{graphicx}
\usepackage{epstopdf}
\usepackage{dcolumn}
\usepackage{bm}
\usepackage{verbatim}
\usepackage[bookmarks=true, bookmarksopen=true, bookmarksnumbered=true, bookmarksopenlevel=1, colorlinks=true, citecolor=blue]{hyperref}
\begin{document}


\title{Measurement of high-k density fluctuation wavenumber spectrum in MAST and Doppler backscattering for spherical tokamaks}

\author{J. C. Hillesheim}
	\email{jon.hillesheim@ccfe.ac.uk}	
	\affiliation{CCFE, Culham Science Centre, Abingdon, Oxon OX14 3DB, United Kingdom}
\author{N. A. Crocker}
	\affiliation{University of California, Los Angeles, Los Angeles, California 90095-7099, USA}
\author{W. A. Peebles}
	\affiliation{University of California, Los Angeles, Los Angeles, California 90095-7099, USA}
\author{H. Meyer}	
	\affiliation{CCFE, Culham Science Centre, Abingdon, Oxon OX14 3DB, United Kingdom}
\author{A. Meakins}	
	\affiliation{CCFE, Culham Science Centre, Abingdon, Oxon OX14 3DB, United Kingdom}
\author{A. R. Field}
	\affiliation{CCFE, Culham Science Centre, Abingdon, Oxon OX14 3DB, United Kingdom}
\author{D. Dunai}
	\affiliation{Wigner Research Centre for Physics, Budapest, Hungary}
\author{M. Carr}
	\affiliation{CCFE, Culham Science Centre, Abingdon, Oxon OX14 3DB, United Kingdom}
\author{N. Hawkes}
	\affiliation{CCFE, Culham Science Centre, Abingdon, Oxon OX14 3DB, United Kingdom}
\author{the MAST Team}
	\affiliation{CCFE, Culham Science Centre, Abingdon, Oxon OX14 3DB, United Kingdom}
	
\date{\today}

\begin{abstract}
The high-k ($7 \lesssim k_{\bot} \rho_i \lesssim 11$) wavenumber spectrum of density fluctuations has been measured for the first time in MAST [B. Lloyd \textit{et al}, Nucl. Fusion \textbf{43}, 1665 (2003)].  This was accomplished with the first implementation of Doppler backscattering (DBS) for core measurements in a spherical tokamak.  DBS has become a well-established and versatile diagnostic technique for the measurement of intermediate-{\it k} ($k_{\bot} \rho_i \sim 1$, and higher) density fluctuations and flows in magnetically confined fusion experiments.  Previous implementations of DBS for core measurements have been in standard, large aspect ratio tokamaks. A novel implementation with 2D steering was necessary to enable DBS measurements in MAST, where the large magnetic field pitch angle presents a challenge.  We report on the scattering considerations and ray tracing calculations used to optimize the design and present data demonstrating measurement capabilities.  Initial results confirm the applicability of the design and implementation approaches, showing the strong dependence of scattering alignment on toroidal launch angle and demonstrating DBS is sensitive to the local magnetic field pitch angle.  We also present comparisons of DBS plasma velocity measurements with charge exchange recombination and beam emission spectroscopy measurements, which show reasonable agreement over most of the minor radius, but imply large poloidal flows approaching the magnetic axis in a discharge with an internal transport barrier.  The 2D steering is shown to enable high-k measurements with DBS, at $k_{\bot}>20\ \mathrm{cm}^{-1}$ ($k_{\bot} \rho_i>10$) for launch frequencies less than 75 GHz; this capability is used to measure the wavenumber spectrum of turbulence and we find $|n(k_{\bot})|^2 \propto k_{\bot}^{- 4.7 \pm 0.2}$ for $k_{\bot} \rho_i \approx 7-11$, which is similar to the expectation for the turbulent kinetic cascade of $|n(k_{\bot})|^2 \propto k_{\bot}^{- 13/3}$.
\end{abstract}

\maketitle
\section{Introduction}

Measurements of fluctuation characteristics and the radial electric field profile are critical for advancing the understanding of a range of phenomena in tokamaks, including turbulent transport, the L-H transition, H-mode pedestal structure, and the effect of applied 3D magnetic field perturbations.  This paper describes the implementation of Doppler Backscattering (DBS) at the Mega Amp Spherical Tokamak (MAST)~\cite{lloyd_overview_2003}.  First the design approach and methodology are described, then the diagnostic implementation is documented.  Experimental data are then used to validate the design calculations and methodology.  With the measurements considered validated, we then apply them to the study of the high-k wavenumber spectrum of density fluctuations.

Doppler backscattering~\cite{hirsch_doppler_2001} (also referred to as Doppler Reflectometry) is essentially a refraction-localized scattering technique, which has been implemented on many fusion experiments~\cite{hennequin_doppler_2004,conway_plasma_2004,hillesheim_multichannel_2009,happel_doppler_2009,tokuzawa_microwave_2012,zhou_microwave_2013}.  For DBS, a millimeter-wave beam is launched into a plasma at a frequency that approaches a cutoff and at an oblique angle to the cutoff surface.  This creates a radially localized region where backscattering occurs off of density fluctuations matching the Bragg condition for $180^{\circ}$ backscattering, $k_{n}=-2 k_i$, where $k_{n}$ is the wavenumber of the scattering density fluctuation and $k_i$ is the incident wavenumber of the diagnostic beam at the scattering location.  The backscattered radiation is then detected at the launch location.  The scattered power is proportional to the density fluctuation power (in the linear scattering regime) and the radiation is Doppler shifted by the lab frame propagation velocity of the turbulent structures.  The Doppler shift is given by $\omega_{DBS}=k_{n}(v_{E \times B} + v_{phase})$, where $v_{E \times B}$ is the equilibrium $E \times B$ drift velocity from the radial electric field and $v_{phase}$ is the phase velocity of the turbulence.

Doppler backscattering has not been implemented before for core measurements in a spherical tokamak.  Measurements at the plasma periphery were reported from Globus-M~\cite{bulanin_observation_2011}.  DBS was implemented at MAST via a temporary transfer of existing microwave hardware~\cite{crocker_high_2011} previously installed on NSTX~\cite{gates_overview_2009}, where it was used for normal-incidence reflectometry.  The transferred hardware consisted of two 8 channel systems that can be used for either conventional reflectometry or Doppler backscattering.  One is a V-band system covering the frequency range 55-75 GHz in 2.5 GHz increments (excluding 65 GHz).  The other is a Q-band system covering 30-50 GHz in 2.5 GHz increments (excluding 40.0 GHz).  The microwave hardware is described in detail in Ref.~\onlinecite{crocker_high_2011} and is also similar to the design in Ref.~\onlinecite{peebles_novel_2010}.

Scattering is an intrinsically three dimensional process, where the vector relation $\mathbf{k_s}=\mathbf{k_i}+\mathbf{k_n}$ must be satisfied to conserve momentum, where the wave-vector indices $s,\ i$, and $n$ refer respectively to the scattered, incident, and density fluctuation waves.  However, the large pitch angle in a spherical tokamaks necessitates that the diagnostic beam be launched at a finite toroidal angle, to match the turbulent fluctuations, which are assumed to be aligned along the field lines ($k_{||}<<k_{\bot}$, where $k_{\bot}$ is the component of the density fluctuation wave-vector perpendicular to both the magnetic flux surface normal and the direction of the magnetic field and $k_{||}$ is the component parallel to the magnetic field).  Differing from implementations in standard aspect ratio devices, independent two-dimensional steering is needed to successfully implement DBS in a spherical tokamak due to the large and variable magnetic field line pitch angle, even for measurements at low wavenumbers.  We present data showing that with a 2D steering capability, DBS can be used to measure high-k, electron-scale density fluctuations ($k_{\bot} \rho_i>10$, where $\rho_i$ is the ion gyroradius).  The detailed discussion of scattering alignment and cross-diagnostic comparisons are necessary to enable the eventual result, where the high-k wavenumber spectrum of turbulence has been measured for the first time in MAST.

After confirming operation of the diagnostic, we present initial physics results.  In particular, measurements indicate poloidal flows within an internal transport barrier with a much larger magnitude than would be expected from existing predictions for neoclassical poloidal rotation.  We also use high-k measurements to present a wavenumber spectrum of density fluctuations at scales far below the ion gyroradius and compare the results to theoretical expectations.  We find that the measured wavenumber spectrum of $|n(k_{\bot})|^2 \propto k_{\bot}^{- 4.7 \pm 0.2}$ for $k_{\bot} \rho_i \approx 7-11$ is not significantly different from the prediction for the kinetic cascade of $|n(k_{\bot})|^2 \propto k_{\bot}^{-13/3}$.  

The paper is organized as follows.  Section~\ref{sec:alignment} uses established scattering theory to arrive at an optimization criterion for DBS alignment.  Pre-installation design considerations and calculations are reported in Sec.~\ref{sec:design}.  The implementation of DBS at MAST using a novel quasi-optical arrangement with 2D steering and a rotatable polarizer can be found in Sec.~\ref{sec:implementation}.  Data analysis methods are briefly discussed in Sec.~\ref{sec:analysis}.  Section~\ref{sec:validate} presents initial data, which is used to demonstrate successful implementation of the diagnostic and to validate the design calculations and methodology.  Cross-diagnostic comparisons of velocity measurements are also reported, with large poloidal flows inferred inside of an internal transport barrier.  Section~\ref{sec:highk} discusses localization of high-k measurements and measurements of the wavenumber spectrum of density fluctuations.  Finally, discussion and conclusions are located in Sec.~\ref{sec:conclusion}.

\section{Wave-vector alignment for DBS \label{sec:alignment}}
In the limit where the electromagnetic wave frequency is much larger than the plasma frequency and electron cyclotron frequency, $\omega >> \omega_{pe},~\omega_{ce}$, and for small fluctuation levels the Born approximation can be used to calculate the scattered electric field for a beam incident on a volume of plasma with density fluctuations~\cite{bekefi_radiation_1966}.  Collective scattering has been investigated in detail~\cite{mazzucato_small-scale_1976,holzhauer_analysis_1978,slusher_study_1980,brower_multichannel_1985,devynck_localized_1993} for measurements of density fluctuations in plasmas and has been employed in several modern experiments~\cite{hennequin_scaling_2004,rhodes_millimeter-wave_2006,mazzucato_short-scale_2008}.  This limit is not well satisfied for DBS, where refraction of the probe beam is fundamental to the technique; however, the same basic concepts apply for wave-vector matching in both techniques.  In addition to refraction deflecting the beam, previous studies of reflectometry have also shown there are distortions to the beam profile near cutoff~\cite{gourdain_assessment_2008,gourdain_application_2008}.  This would be expected to affect calculation of the spectral resolution for DBS.  These considerations limit the accuracy of simplified analytical results.  The following should be sufficiently accurate for optimization of design, where results of the calculations can then be validated experimentally.

It can be shown that the electric field scattered by plasma density fluctuations follows~\cite{bekefi_radiation_1966,slusher_study_1980}
\begin{widetext}
\begin{align} \label{eqn:es}
{\rm \bf E_s}(\omega_s)=& \left[ \hat{\rm \bf k}_s \times  \left( \hat{\rm \bf k}_s \times {\rm \bf E_0} \right)  \right]
\frac{ r_0}{16 \pi^4 \sqrt{a_x a_y}} \int  d \omega_n  dt d{\rm \bf k}_n d {\rm \bf x}   \\ \nonumber
& \widetilde{n}({\rm \bf k_n}, \omega_n) 
e^{-x^2/a_x^2} e^{-y^2/a_y^2} 
e^{-i \left( \omega_i+\omega_n-\omega_s \right)t}
e^{i\left({\rm \bf k}_i+{\rm \bf k}_n-{\rm \bf k}_s \right) \cdot {\rm \bf x}} e^{-i\omega_s R_d/c} +{\rm c.c.}
\end{align}
\end{widetext}
Here $x$ and $y$ are the directions transverse to the propagation of the beam and $z$ is along the axis of the beam, while ${\rm \bf x}$ is the position vector.  $E_0$ is the incident electric field and $\widetilde{n}({\rm \bf k_n}, \omega_n)$ is the spectrum of density fluctuations.  We have assumed a Gaussian beam with beam widths $a_x$ and $a_y$, which vary along $z$.  The classical electron radius is $r_0=e^2/4\pi\epsilon_0m_ec^2$.  The indices $j=n,i,s$ for the wave-vector and frequency are for the density fluctuations, incident radiation, and scattered radiation, respectively; $R_d$ is the distance the scattered radiation travels to the detector.

Integration over time and space result in the selection rules ${\rm \bf k}_i+{\rm \bf k}_n={\rm \bf k}_s$ and $\omega_i+\omega_n=\omega_s$ and Eqn.~\ref{eqn:es} can be reduced to an expression for the dependence of the intensity of the scattered power along the beam path due to alignment between the density fluctuation and scattered wave:
\begin{widetext}
\begin{equation} \label{eqn:di}  
dI \propto dz ~ \widetilde{n}^2({\rm \bf k_n}, z) 
\exp \left( -\frac{(k_{n,x}-k_{s,x})^2 a_x^2}{2} \right)
\exp \left( -\frac{(k_{n,y}-k_{s,y})^2 a_y^2}{2} \right),
\end{equation}
\end{widetext}
where $k_{j,l}$ are the wave-vector components.  The Gaussian factors arise due to the assumption of a Gaussian incident beam. To apply this to the DBS technique, we first assume that the scattering is highly localized to the turning point (minimum perpendicular index of refraction along the path) of the beam from the increase in the amplitude of the electric field and decrease in wavenumber of the beam due to refraction.  At the turning point, where $k_{i,y}=k_{s,y}=0$, scattered power will be maximized for $k_{n,y}=0$ (\textit{i.e.} $k_r=0$ in the usual notation, where the radial direction is normal to the flux surface), so the second exponential becomes unity.  For simplicity, take $a_x=a_y=a_0$.  Define $\theta_{mis}$ as the mismatch angle, $\hat{\rm \bf k}_i \cdot \hat{\rm \bf B}=\cos (\pi/2-\theta_{mis})$, so that scattered power is maximal for $\theta_{mis}=0$ (assuming a monostatic antenna arrangement). We are interested in the direct backscattered light that returns to the detector, $k_{s,x}=0$.  In the incident beam frame, $k_{n,x}=|{\rm \bf k}_n| \sin \theta_{mis}$.  Equation~\ref{eqn:di} is then
\begin{equation} \label{eqn:exp}
dI \propto dz ~ \widetilde{n}^2({\rm \bf k_n}, z) 
\exp \left( -\frac{|{\rm \bf k}_n|^2 a_0^2 \sin^2 \theta_{mis}}{2} \right).
\end{equation}
In terms of the scattering wavenumber of density fluctuations and for small $\theta_{mis}$, we arrive at the criterion that for significant scattered power (taken for convenience to be $1/e$):
\begin{equation} \label{eqn:rule}
|\theta_{mis}| \lesssim \frac{ \sqrt{2}}{k_n a_0}.
\end{equation}
Although some effects are not included, such as the cutoff surface and beam curvatures, this provides a rule-of-thumb for design considerations.

Another criterion for optimizing DBS alignment that has been used is $| k_{||}/k_{\bot} |<0.1$~\cite{conway_irw_2011} at the ray turning point.  Both expressions are optimized at the same condition since $k_{||}=0$ at $\theta_{mis}=0$.  However, since $ k_{||}/k_{\bot}=\tan \theta_{mis}$ (at the turning point, $k_r=0$), the criterion from Ref.~\onlinecite{conway_irw_2011} is equivalent to $|\theta_{mis}| \lesssim 6^{\circ}$, which lacks the dependence on wavenumber and beam size.  For illustration, taking typical DIII-D parameters (for which published measurements of beam size exist), DBS hardware and quasi-optical systems~\cite{hillesheim_multichannel_2009,peebles_novel_2010,rhodes_quasioptical_2010} result in $a_0 \approx 2.5 \ {\rm cm}$ (measured in vacuum) and steerable mirrors allow access to $k_{n,\bot} \approx 4$ to $16 \ {\rm cm}^{-1}$.  Using these parameters in Eqn.~\ref{eqn:di} yields $|\theta_{mis}| \lesssim 8^{\circ}$ and $2^{\circ}$, respectively.  Note that the restriction at high wavenumbers could be even more constraining if refractive effects elongating the beam~\cite{gourdain_assessment_2008,gourdain_application_2008} are taken into account.  For low wavenumbers in a moderate to large aspect ratio tokamak there is a weak constraint on toroidal alignment from pitch angle mismatch and toroidal steering only has a significant effect on high-k measurements.  In a spherical tokamak, where the pitch angle of the magnetic field can be $\sim 35^{\circ}$, two dimensional steering is \textit{necessary}, even for low-k DBS measurements.

It is also worth noting that a finite $k_{||}$ for the beam should not be associated with scattering from the parallel component of the fluctuation, which should be comparable to the field line connection length, $k_{{n},||} \sim 1/qR$, and very small compared to the other components.  Using notation defined by the magnetic field, for significant misalignment ($k_{n,||} << k_{i,||}, k_{s,||}$) the selection rule for wave-vector gives $k_{i,||}+k_{n,||}=k_{s,||} \rightarrow k_{i,||} \approx k_{s,||}$.  To conserve momentum, the scattered beam is effectively mirrored with respect to the field line.  It is important to note that this occurs due to the separation of scales between the beam spot size and the fluctuation parallel wavelength -- for significant misalignment there is no $k_{{n},||}$ on the scale of the beam, so the beam $k_{||}$ is conserved.  For a fixed monostatic antenna, this means the effect of finite $\theta_{mis}$ is that the center of the scattered beam is re-directed away from the antenna position, resulting in reduced detected power (assuming the incident beam is not detected at all).  The Doppler shift of the detected radiation will still be dominated by $k_{n,\bot}$, regardless of the misalignment, which is shown later with experimental data.

For significant misalignment, it might be possible for the exponential dependence of the scattering alignment to overcome the (often assumed) power law dependence of the density fluctuations, which is the primary localization mechanism (when combined with the change in wavenumber of the beam due to refraction along the beam path).  This could result in a poorly localized signal with comparable scattered power from a large radial region of the plasma.

\section{DBS design for MAST \label{sec:design}}
This section describes design of a Doppler backscattering implementation for MAST.  Doppler backscattering has not been implemented before for core measurements in a spherical tokamak.  Measurements at the plasma periphery were reported from Globus-M~\cite{bulanin_observation_2011}.  In this section we discuss a number of factors impacting implementation of the diagnostic.  In principle, all of these considerations also matter for implementation in standard aspect ratio devices, but the implementation for a spherical tokamak is more challenging, as discussed below.

\begin{figure}[!htbp]
\includegraphics[width=8.5 cm]{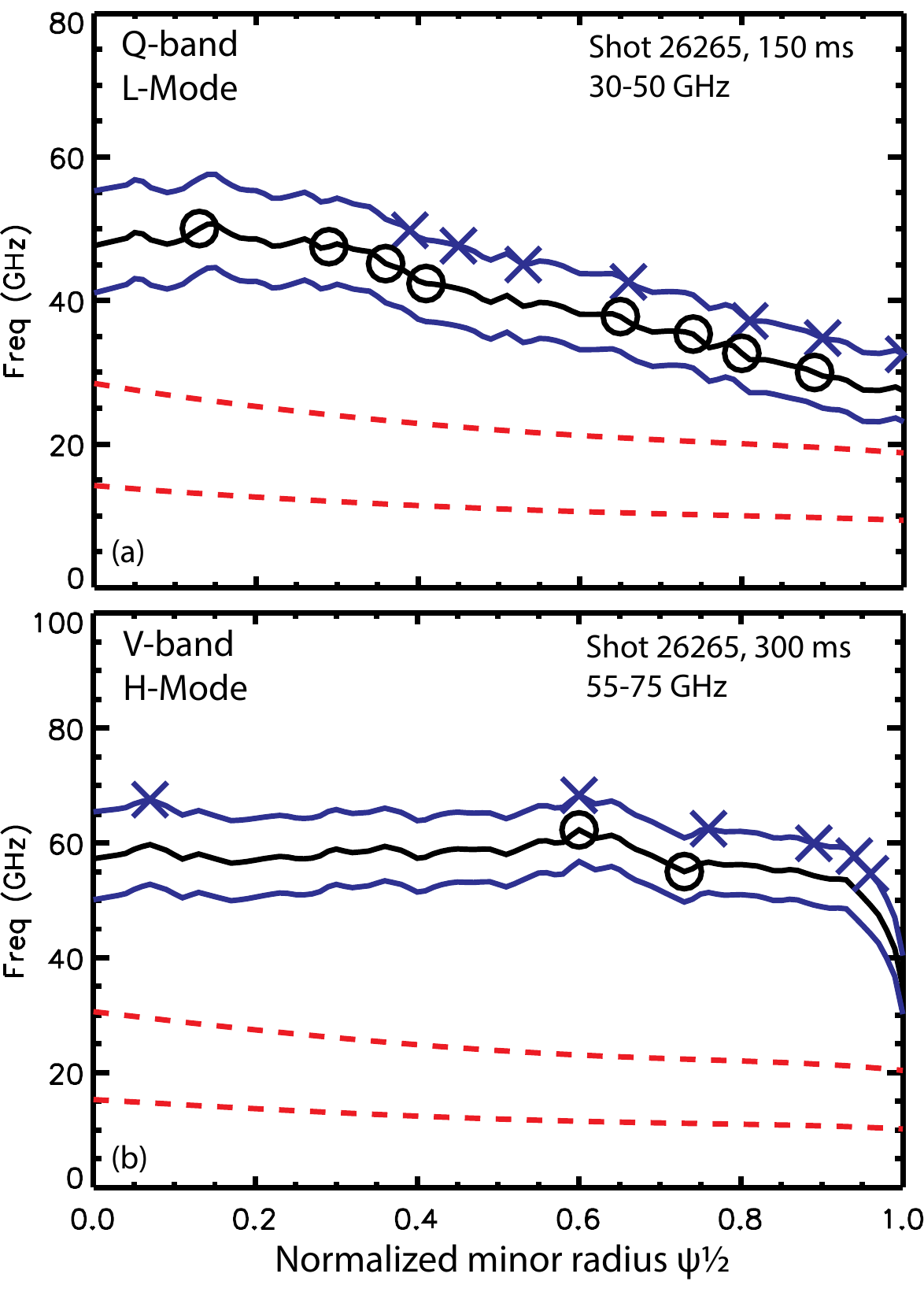}
\caption{\label{fig:mast_cutoffs} Indicative normal incidence cutoffs and resonances for two times in a MAST shot: (a) lower density L-mode and (b) H-mode.  Red dashed lines are the first and second electron cyclotron harmonics.  Solid black lines are O-mode cutoffs.  Solid blue lines are X-mode cutoffs.  For frequencies launched by the (a) Q-band and (b) V-band system, X's indicate location of X-mode cutoffs and O's indicate location of O-mode cutoffs.  Symbols for frequencies that would shine through placed at maximum cutoff frequency; path effects not accounted for.  Only cutoffs corresponding to one 8-channel system or the other are shown in each panel, for clarity.}
\end{figure}

Since the systems were originally designed for NSTX, there is similar density profile access in MAST.  Figure~\ref{fig:mast_cutoffs} shows an overview of profile access during a MAST shot, where the first two cyclotron resonances, and O- and X-mode cutoffs are plotted.  Symbols indicate location of cutoffs for frequencies launched by the two systems.  During low density L-mode periods, the Q-band system provides coverage of most of the inner radius, typically covering from near the edge to $\sqrt{\psi} \sim 0.5$ (where $\sqrt{\psi}$ is the normalized square root of the poloidal flux).  This is ideal for core turbulence studies when used for either reflectometry or DBS, and for L-H transition studies.  The V-band system is expected to either not encounter a cutoff or encounter one deep in the core plasma in L-mode.  In H-mode plasmas, the Q-band system accesses the lower two-thirds to half of the pedestal, while the V-band system accesses the top of the pedestal, and possibly core locations, depending on details of the density profile and which polarization is used.

\subsection{Ray tracing for scattering alignment \label{sec:rays}}

\begin{figure}[!htbp]
\includegraphics[width=8.5 cm]{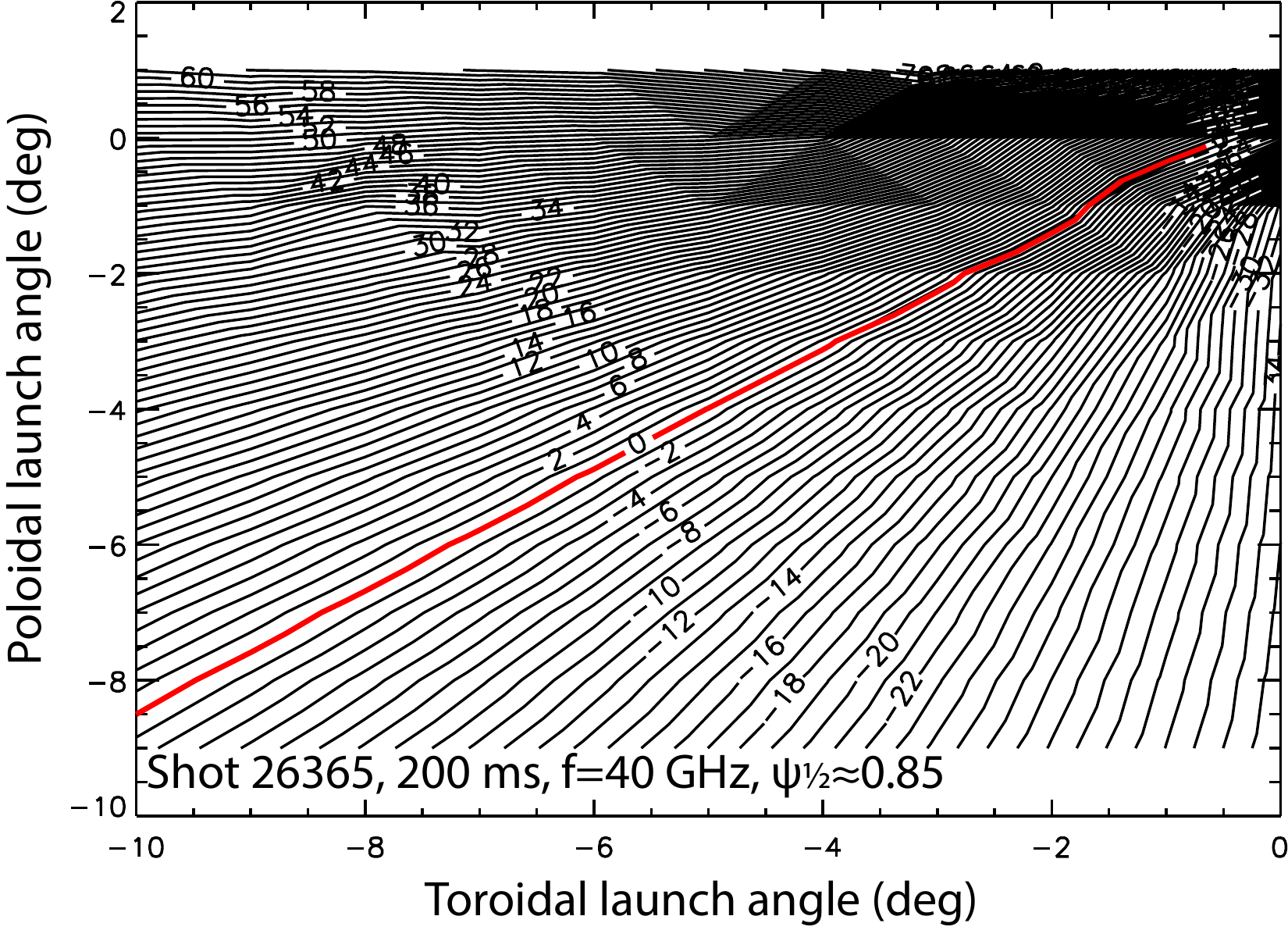}
\caption{\label{fig:theta_mis_single} Contours of $\theta_{mis}$ at the ray turning point as a function of poloidal and toroidal launch angle.  Launch frequency is 40 GHz in X-mode during an L-mode portion of a MAST shot.  The optimal condition of $\theta_{mis}=0^{\circ}$ is plotted in red.}
\end{figure}

Ray tracing relying on the Genray code~\cite{smirnov_genray_1995} has been used for MAST, with experimental magnetic equilibria from EFIT~\cite{lao_reconstruction_1985,appel_efitpp_1995} and density profiles from a 130 point Thomson scattering system~\cite{scannell_130_2010}.  The Appleton-Hartree (cold plasma) dispersion relation is used.  Ray tracing is used to calculate the $\theta_{mis}$ parameter (via the expression $\hat{\rm \bf k}_i \cdot \hat{\rm \bf B}=\cos \left(\pi/2-\theta_{mis} \right)$) to assess several design and implementation considerations.  Figure~\ref{fig:theta_mis_single} shows the result of calculating $\theta_{mis}$ at the ray turning point (minimum perpendicular index of refraction along the ray) as a function of toroidal and poloidal launch angle, originating from the MAST port window used for the DBS implementation.  The plasma chosen in this example is in L-mode and the launch frequency of 40 GHz in X-mode polarization approaches cutoff at $\sqrt{\psi} \approx 0.85$.  The optimal launch condition of $\theta_{mis}=0^{\circ}$ is plotted in red.  Although the corresponding wavenumbers are not included in the plot, the calculations projected that about $10^{\circ}$ 2D steering would enable measurements of $k_{{n},\bot}\sim 4-12 \ {\rm cm}^{-1}$.  With the relatively low magnetic field in MAST, this corresponds to $k_{{n},\bot} \rho_i \sim 2-7$.  One can also see from the plot that accurate beam steering is essential.  Using Eqn.~\ref{eqn:rule} for an estimate and assuming $a_0 \approx 4$ cm (which is similar to measured vacuum values), one finds for higher wavenumbers at fixed poloidal angle, the toroidal launch angle must be accurate to within $\sim \pm 1^{\circ}$, while for lower wavenumbers the launch must still be accurate to within $\sim \pm 2^{\circ}$.  These pre-installation calculations ended up being roughly consistent with actual measurements, as shown later. 

\begin{figure}[!htbp]
\includegraphics[width=6.0 cm]{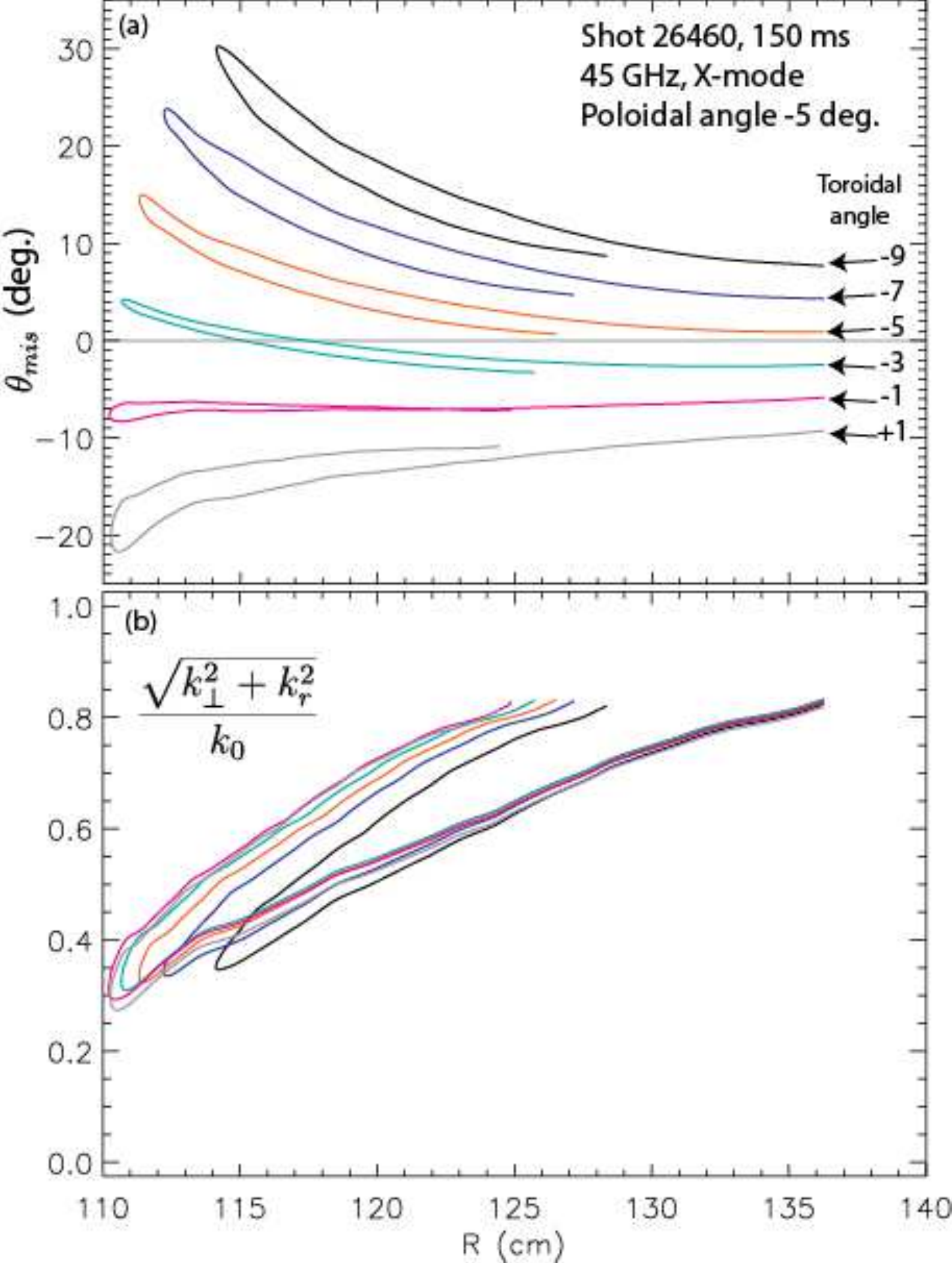}
\caption{\label{fig:theta_path} Ray tracing results showing (a) misalignment angle $\theta_{mis}$ and (b) relative perpendicular wavenumber of the beam versus major radius, along the ray path as a function of toroidal angle in an L-mode MAST plasma.  Scattering wavenumber is $k_{n,\bot} \sim 6 \ {\rm cm}^{-1}$ and location is $\sqrt{\psi} \approx 0.5$.  Arrows in (a) indicate toroidal launch angle for rays, plotted with solid lines.}
\end{figure}

Figure~\ref{fig:theta_path} compares the misalignment angle along the ray path for different toroidal launch angles at a fixed poloidal angle.  Figure~\ref{fig:theta_path}(a) plots the misalignment parameter $\theta_{mis}$, which would reduce the detected power, and Fig.~\ref{fig:theta_path}(b) shows the change in perpendicular wavenumber along the path, which is the dominant localization mechanism.  Note here that since we are considering alignment along the entire beam path and not local to the turning point, it is the total perpendicular wavenumber, $\sqrt{k_{\bot}^2+k_{r}^2}$, that enters into the determination of $\theta_{mis}$.  The case chosen is from an L-mode time period where the launch frequency of 45 GHZ would be reflecting near $\sqrt{\psi} \approx 0.5$ and scattering from plasma fluctuations with $k_{n,\bot} \sim 6 \ {\rm cm}^{-1}$ with the poloidal launch angle of $-5^{\circ}$.  The plots are as a function of major radius, which is the reason for the different ending positions of the rays (i.e. they all end at the last closed flux surface, but at different vertical locations resulting from the toroidal and poloidal drift of the beam).  The optimal toroidal launch angle would be between $-2^{\circ}$ and $-3^{\circ}$.  It is interesting to note that $|\theta_{mis}|$ is at a maximum in most cases at the minimum $\sqrt{k_{\bot}^2+k_{r}^2}/k_0$; that is, the misalignment is worst nearest the cutoff.  This makes sense, as the ray is traveling close to normal to the flux surfaces for much of the ray path, so $\theta_{mis}$ does not depend strongly on the pitch angle away from cutoff.  The misalignment between the ray and field lines manifests itself when the ray is traveling tangential to the flux surfaces, at the turning point.  Stated another way: refraction mostly (exactly so in a slab) changes the component of the wave-vector in the direction of the index of refraction gradient, so $k_{||}$ is approximately conserved (\textit{i.e.} it changes little compared to the change in $k_r$).  Therefore $k_{||}/k_0$ is typically at a maximum closest to the cutoff.  

Figure~\ref{fig:theta_path} illustrates several points.  One is that rays that are misaligned near the cutoff can cross $\theta_{mis}=0^{\circ}$ along the ray path, which could provide a degree of localization for backscattering along the beam path.  A second point is that for rays with significant misalignment, the reduced power from $|\theta_{mis}|$ is opposing the localization by $\sqrt{k_{\bot}^2+k_{r}^2}/k_0$.  Depending on the magnitude of the misalignment, how the beam size varies along the path, and the wavenumber spectrum of the turbulence, this could plausibly impact measurement localization.  However, due to the exponential dependence of the scattered power on $\theta_{mis}$, this should not impact well-aligned measurements. It is also notable in Fig.~\ref{fig:theta_path}(b) that toroidal misalignment has relatively little impact on $k_{n,\bot}$, which is determined dominantly by the poloidal launch angle.  This is due to the same argument as for $k_{||}$ conservation:  the index of refraction gradient is mostly in the radial direction, so $k_{\bot}$ is mostly conserved. 

\begin{figure}[!htbp]
\includegraphics[width=8.5 cm]{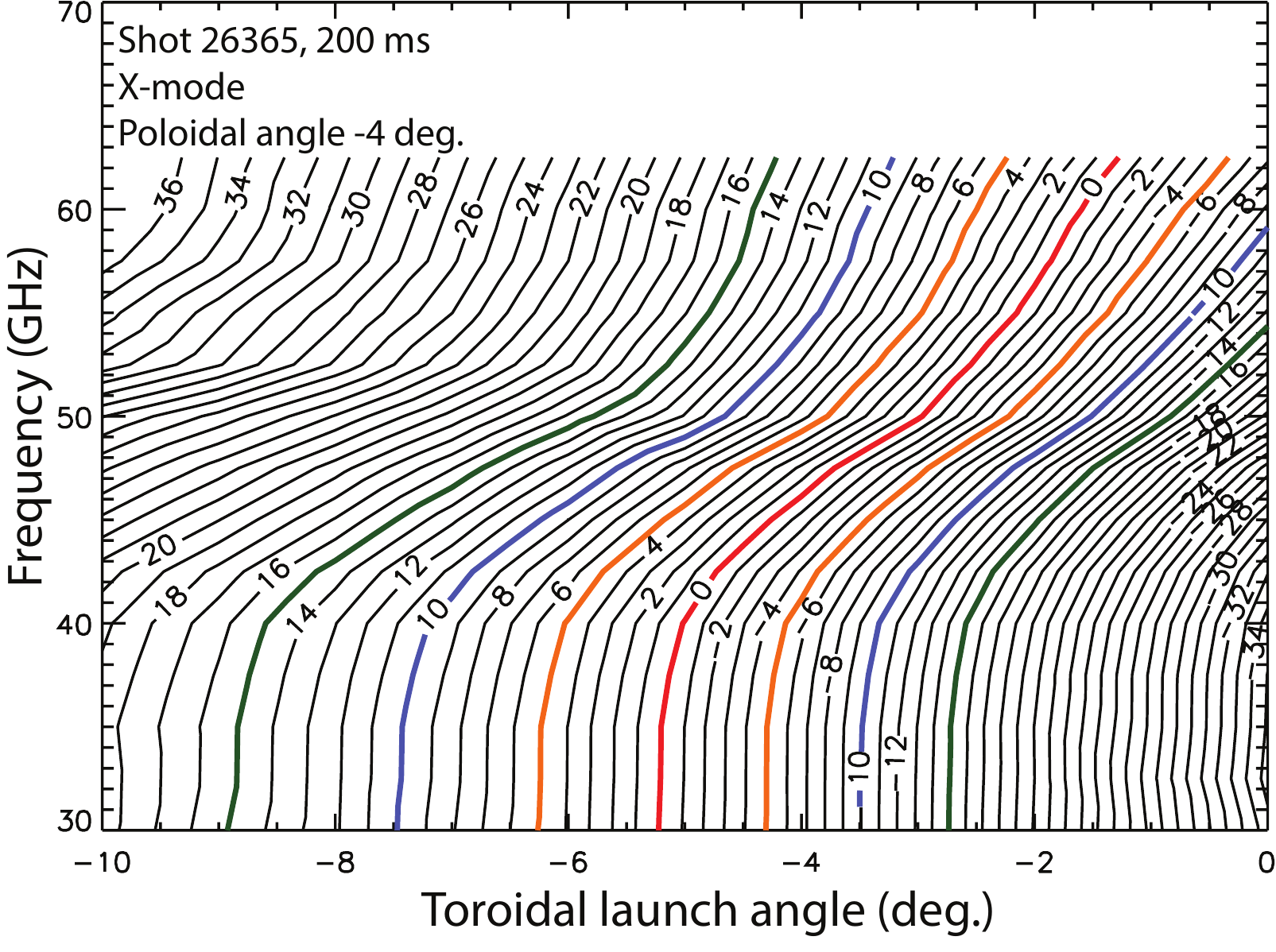}
\caption{\label{fig:theta_freq} Ray tracing results showing contours of misalignment angle $\theta_{mis}$ (at the ray turning point) as a function of frequency and toroidal launch angle for an L-mode MAST plasma at a fixed poloidal angle of $-4^{\circ}$.  The 30 GHz cutoff is close to the last close flux surface and the 62.5 GHz cutoff would be at $\sqrt{\psi} \approx 0.3$.  The fluctuation wavenumbers would range from about 3 to about 10 cm$^{-1}$, respectively.  Select contours colored for ease of viewing.}
\end{figure}

An example of the dependence of $\theta_{mis}$ at the ray turning point on launch frequency and toroidal launch angle is illustrated in Fig.~\ref{fig:theta_freq}.  This calculation is of particular interest since the 16 launched frequencies would effectively be a vertical cut through such a plot (with some differences due to polarization), for a particular launch direction.  For frequencies of 30, 40, 50, and 60 GHz a poloidal launch angle $-4^{\circ}$ in X-mode, the respective scattering locations would be $\sqrt{\psi}\sim$0.95, 0.85, 0.55, and 0.30, while the scattering wavenumbers would be $k_{n,\bot} \sim$ 3, 4, 6, and 10 cm$^{-1}$.  The allowable misalignment from Eqn.~\ref{eqn:rule}, assuming $a_0 \approx 4$ cm, would then be about 7$^{\circ}$, 5$^{\circ}$, 3$^{\circ}$, and 2$^{\circ}$.  Due to changes in pitch angle and narrowing of the allowable mismatch, the number of channels that can be simultaneously aligned was projected to be limited.  The calculations did project that in many cases 1/3 to 1/2 of the minor radius would simultaneously be within reasonable alignment for L-mode portions of shots.  Pedestal measurements in H-mode plasmas should not be greatly impacted, due to their close spatial proximity.  

\begin{figure}[!htbp]
\includegraphics[width=6.0 cm]{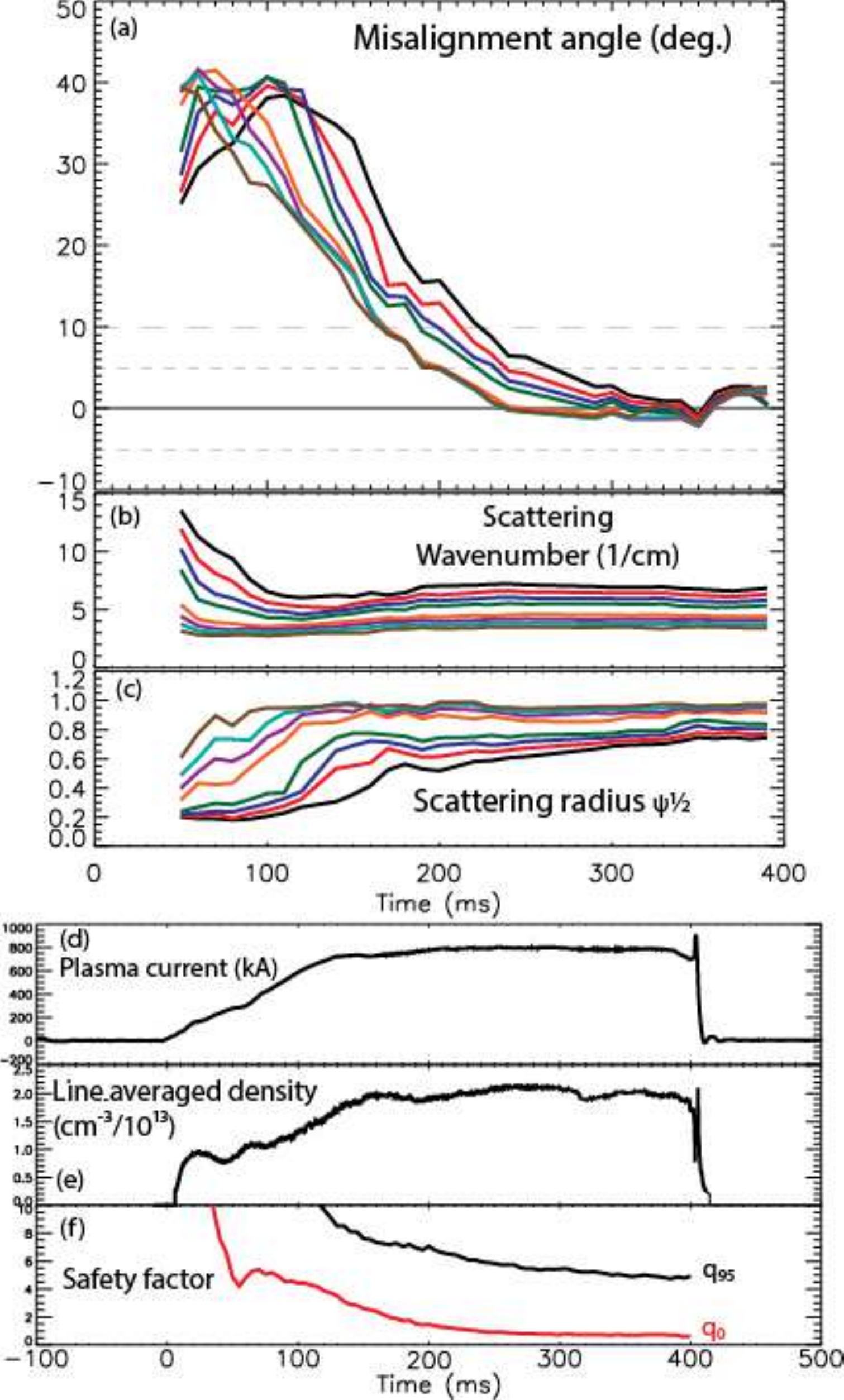}
\caption{\label{fig:theta_time} Ray tracing results showing from a poloidal launch angle of $-4^{\circ}$ and toroidal launch angle of $-6^{\circ}$ in an L-mode MAST plasma as a function of time showing (a) the misalignment angle at the turning point, (b) scattering wavenumber, and (c) scattering normalized minor radius.  Q-band system frequencies, 30-50 GHz,  launched in X-mode.  Also plotted are the (d) plasma current, (e) line averaged density, (f) and safety factor at the edge and on axis.}
\end{figure}

An additional challenge for implementation in MAST is the relatively short plasma duration, $\sim 0.5$ s.  Diffusion does not typically have sufficient time to relax the current profile to a steady-state (unless, for instance, the current profile is dominated by the bootstrap current).  This results in a continuously evolving safety factor profile, which directly impacts magnetic field pitch angle and therefore scattering alignment.  Figure~\ref{fig:theta_time} shows the evolution of misalignment angle at the turning point over time, along with the scattering wavenumber and location.  Also plotted are the plasma current, density, and safety factor at the edge and on axis.  Horizontal lines added to Fig.~\ref{fig:theta_time}(a) for reference at $0^{\circ}$, $\pm 5^{\circ}$, and $10^{\circ}$.  One can see that there is over 150 ms time period where all frequencies are aligned to within $|\theta_{mis}| < 10^{\circ}$, with over 100 ms where $|\theta_{mis}| < 5^{\circ}$ for all channels.  This shows that the outer $\sim 1/3$ of the plasma should be able to be accessed simultaneously towards the end of the discharge.  The effect of scanning toroidal launch angle is essentially to move results as in Fig.~\ref{fig:theta_time}(a) up or down in $\theta_{mis}$.  From Fig.~\ref{fig:theta_time}(a) we expect no localized DBS signal early in the shot, but for the signal to appear around 150 ms if alignment is chosen appropriately for the current flat top.

A final issue to be considered is non-WKB effects related to polarization interaction in steep density gradients.  The ray tracing calculations presented here assume a WKB or geometrical optics approach is valid, and that O-mode and X-mode can be treated as distinct normal modes of wave propagation with negligible interaction.  It is possible for interaction between the normal modes when there is large magnetic shear or the density gradient is large~\cite{ginzburg_propagation_1961,budden_theory_1952,zheleznyakov_linear_1983}.  This effect is significant when the difference between the O-mode and X-mode wavenumbers small, $|k_{\mathrm{O}} - k_{\mathrm{X}}| << 2\pi/L$, where $L$ is the plasma inhomogeneity length scale.  Due to the low magnetic field in a spherical tokamak the left hand side can be small.  Although there can be large global magnetic shear (a large gradient of $q(\psi)$), most of the local shear is located on the high field side due to the large gradient in toroidal magnetic field from the $1/R$ dependence near the center column.  For the typical measurement region with DBS, on the low field side of the tokamak, there is slow variation of the magnetic field pitch angle, so the dominant inhomogeneity affecting microwave propagation is due to the density profile.  In the H-mode pedestal, the density gradient scale length can be on the order of 1 cm.  However, in the region where scattering is localized for DBS, either $k_{\mathrm{O}}$ or $k_{\mathrm{X}}$ should be small due to approaching one of the cutoff surfaces or the other, depending on the launch polarization, so at least for the region where the scattering is localized the criterion for significant interaction will rarely be satisfied.  The exception would be when the H-mode pedestal height is large enough to contain both cutoffs, in which case the interpretation of measurements becomes more complicated.  For the high-k measurements discussed in Sec.~\ref{sec:highk}, there would also be less difference between the wavenumbers, but there are usually not large density gradients in the core.  For measurements localized in the core of an H-mode plasma $k_{\mathrm{O}}$ and $k_{\mathrm{X}}$ could be similar when the beam propagates through the pedestal.  The result of polarization interaction somewhere along the beam path, but not in the DBS localization region near cutoff, should be indistinguishable from misalignment between the launched polarization and the magnetic field pitch angle at the edge, which can result in detected Doppler shifts from both the X-mode and O-mode cutoffs.

\section{DBS Implementation \label{sec:implementation}}
This section describes the MAST DBS implementation, in particular the novel quasi-optical arrangement with a 2D steering mirror and a rotatable polarizer that is used to combine the beams of the two microwave systems.  Due to the possible sensitivity of measurement interpretation to aspects of these components, detailed measurements characterizing them are presented.  Data from the systems was acquired during experiments at 10 MHz with 14 bit resolution; typically only 15 channels were digitized, with the 75 GHz channel replaced with a timing reference.  The systems were operated in a monostatic antenna arrangement, which results in DC offsets due to internal reflections.  Amplifiers with adjustable DC offsets, for each of the 32 data channels, were used to compensate for this effect.  The dynamic range between system noise levels and amplifier saturation was about 30 dB in amplitude (60 dB in power), although core measurements rarely used the full range.



\subsection{Quasi-optical system}

\begin{figure}[!htbp]
\includegraphics[width=14 cm]{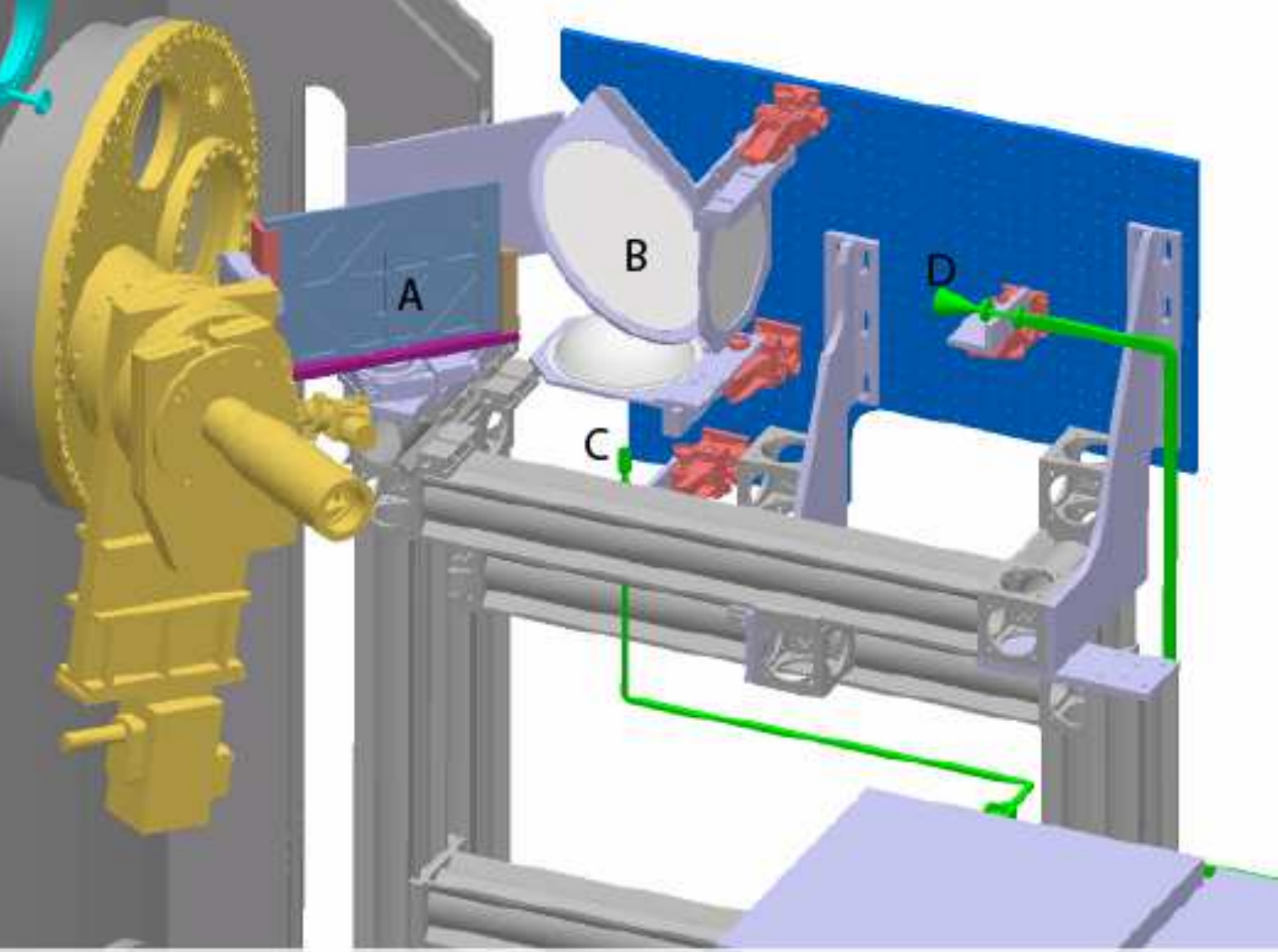}
\caption{\label{fig:drawing}  Computer rendering of quasi-optical system beam system and support frame installed at MAST.  Labeled components: (A) steering mirror, (B) rotatable polarizer, (C) V-band antenna, and (D) Q-band antenna.}
\end{figure}

A quasi-optical Gaussian beam system was designed to implement the two 8-channel millimeter-wave diagnostic systems for MAST within available space and port access constraints.  Each system was operated monostatically, in orthogonal polarizations, with a scalar (V-band) or conical (Q-band) horn feeding an aspherical high density polyethylene lens.  The horns are located slightly separated from the lens focal lengths to image the feed antennae for optimal beam waist size and location as determined by laboratory tests.  After the lenses, the beams are combined via a rotatable polarizer, which can be adjusted to match the magnetic field pitch angle so each system ideally operates in one linear polarization within the plasma.  The combined beamline is then reflected by two mirrors, the second of which is remotely steerable in two dimensions for scattering alignment.  The mirror was used for steering on a shot-to-shot basis, while the rotatable polarizer (to match the magnetic field pitch angle) and waveguide twists (to change which system operated in which polarization) were adjusted day-to-day as determined by experimental objectives.  Short fundamental waveguide lengths, about 1 m in length, were used to connect the horn antennas to the microwave hardware.  Directional couplers connect the launch and receive waveguide for monostatic operation.  Figure~\ref{fig:drawing} shows a computer rendering of the quasi-optical system used at MAST, with some detail omitted (\textit{e.g.} absorbent materials were added in many places on the frame).  The final steering mirror is remotely controlled via the rotation stages that can be seen below and to the left of the mirror in the drawing.  References to the toroidal launch angle refer to the angle about the mirror axis as viewed from above, with positive being counter-clockwise.  The steering mirror center, where the toroidal and poloidal steering axes intersect, is about 1 m from the last closed flux surface and 2.4 m from the center of the torus and on the machine mid-plane.  The port window used was offset from the center of the large pictured flange (the normal to the center of which was parallel to the major radius) by 12.5 cm, allowing a larger effective toroidal angle with the plasma in one direction than the other.


\subsection{Gaussian beam characterization}

\begin{figure}[!htbp]
\includegraphics[width=8.5 cm]{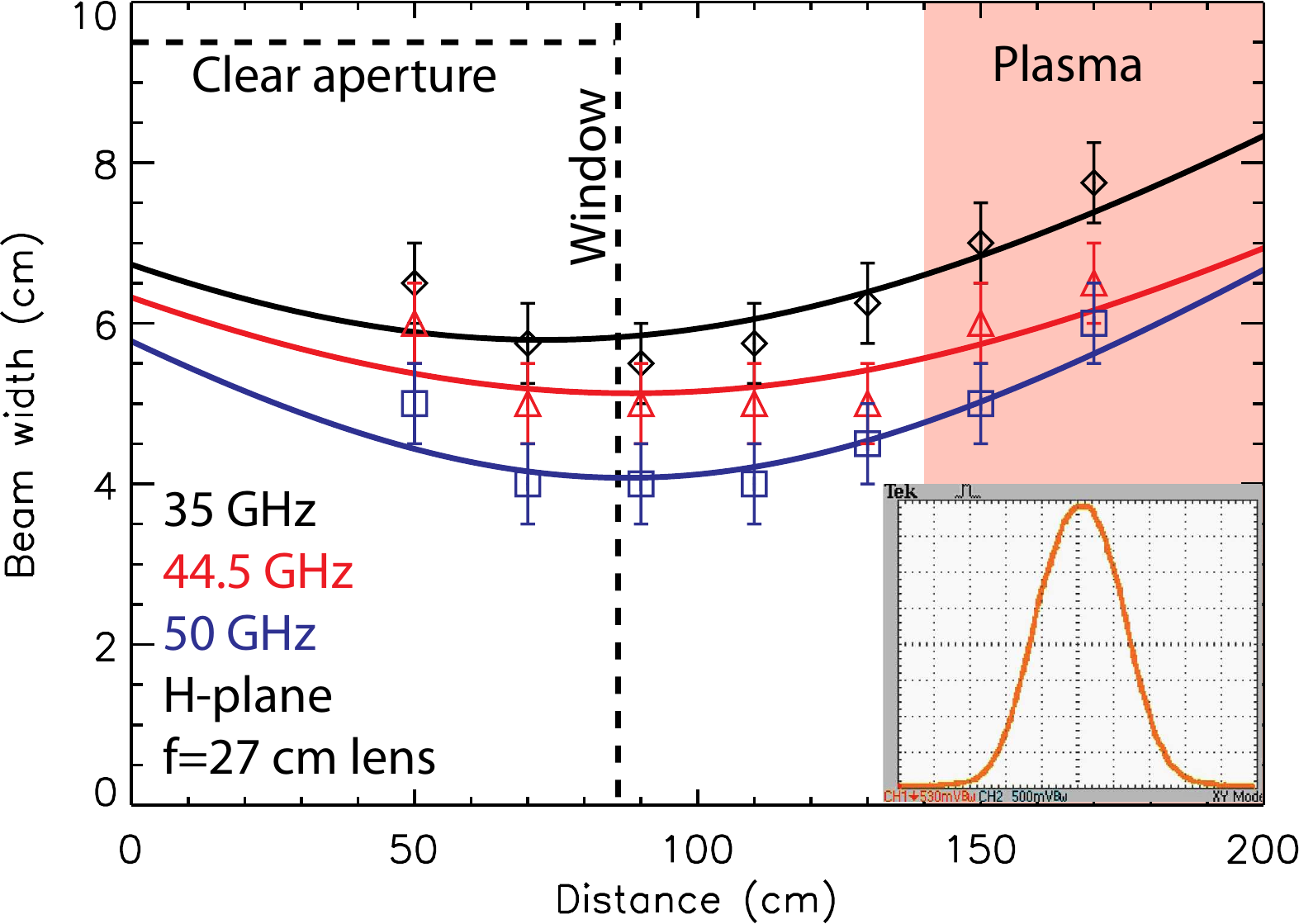}
\caption{\label{fig:qband_beamprofs}  Laboratory measurements of beam radius profiles at Q-band frequencies.  Solid lines are fits to expected beam size variation for a Gaussian beam.  The window location, window clear aperture, and approximate plasma location are annotated.  Inset is the measured beam H-plane intensity profile for 50 GHz at 130 cm, with a horizontal scale of 2 cm/div.}
\end{figure}

The design of the system was limited by pre-existing equipment.  The placement of the steering mirror and size of the beam at the window are the resulting limitations on steering angle.  Figure~\ref{fig:qband_beamprofs} shows laboratory measurements of beam profiles at three frequencies with the conical horn used in the MAST DBS implementation.  The H-plane measurements were found to be wider than the E-plane.  Distance in the figure is referenced to the lens location and taken along the beam path.  Shown inset is the measured H-plane beam intensity profile for 50 GHz at 130 cm; horizontal scale is 2 cm/division.  Data for each frequency were fit to the expression for the expected radius of a Gaussian beam, $w(z)=w_0\sqrt{1+((z-z_0)/z_r)^2}$, where $w_0$ is the 1/$e^2$ intensity radius at the beam waist, $z_0$ is the location of the beam waist, and $z_r=\pi w_0^2/\lambda_0$ is the Rayleigh range, $\lambda_0$ is the vacuum wavelength.  Although the window coincides with the beam Rayleigh range, the beam size at the low frequencies of the Q-band system was expected to limit the steering range to $\sim \pm 7^{\circ}$.  \textit{In situ} tests later confirmed this limitation.

\subsubsection{Wavenumber resolution}
The design of a DBS system should ideally include optimization of the wavenumber resolution of the diagnostic and beam size in the plasma, see for instance Refs~\onlinecite{hirsch_doppler_2004,happel_doppler_2009,rhodes_quasioptical_2010}.  Due to limitations on the MAST DBS implementation, in vessel components were not possible and the external quasi-optical system described above was necessary.  The beam was optimized such that the beam waist was close to the vacuum window, enabling the largest possible angular range for steering.  Here we assess the impact of that choice on the wavenumber resolution of the diagnostic.  An estimate for the wavenumber resolution for DBS measurements is given by~\cite{hirsch_doppler_2001}
\begin{equation}
\Delta k = \frac{2 \sqrt{2}}{w} \sqrt{1 + \left( \frac{w^2 k_0}{ R_c} \right)^2 },
\end{equation}
which is for a Gaussian beam with amplitude profile $E \propto ^{-r^2/w^2}$; $k_0$ is the vacuum wavenumber and $R_c$ is the effective radius of curvature given by $R_c={R_{beam}R_{cutoff}/(R_{beam} + R_{cutoff})}$, and $R_{beam}$ and $R_{cutoff}$ are respectively the beam and cutoff layer radii of curvature.  Using the measured beam profiles in Fig.~\ref{fig:qband_beamprofs} and taking $R_{cutoff} =60$ cm near to the plasma edge yields a range $2.2 \ \mathrm{cm}^{-1} \lesssim \Delta k \lesssim 3.3 \ \mathrm{cm}^{-1}$ for the Q-band system frequencies.  For the range of accessible wavenumbers, this corresponds to $\Delta k/k_{\bot} \approx 0.3-0.5$; this is larger than would be optimal, but still suitable for measurements.  This relatively large $\Delta k/k_{\bot}$ results in the broad frequency peak shown in measurements below with the Q-band system.  Note that $R_{cutoff}=60$ roughly approximates an O-mode cutoff surface near the edge, but the radius of curvature for X-mode would be larger, reducing $\Delta k$.  Since the V-band system usually accessed the core and measured higher $k_{\bot}$ due to viewing geometry, $\Delta k/k_{\bot}$ was smaller.  

\subsection{Rotatable polarizer alignment}
A 12 inch circular rotatable polarizer is used to combine the two systems and to match the beam polarizations to the magnetic field pitch angle at the edge of the plasma.  The polarizer was custom-made and constructed of copper lines etched into a substrate material.  Waveguide twists are used so that each antenna launches at 45$^{\circ}$ from vertical.  Each waveguide run also has a 90$^{\circ}$ twist and a replacement straight piece of waveguide of the same length, so that each system can be used in either X-mode or O-mode polarization.  The magnetic field pitch angle at the edge in MAST is approximately $30-35^{\circ}$ (depending on plasma current), so a small amount of power is lost from reflection with the antennas at $45^{\circ}$.

The polarizer itself is at a $45^{\circ}$ angle from vertical, which must be taken into account for polarization alignment with the field.  It is necessary to know what angle the polarizer should be set to in order to match a given magnetic field pitch angle.  The polarizer rotation angle is $\theta_{rot}$, with 0$^{\circ}$ corresponding to the wires oriented vertically in the MAST scheme.  Let $\theta_{p}$ be the magnetic field pitch angle, $\theta_{l,i}$ (with $i=v,q$ the system) be the polarization angle of the launched radiation transmitted/reflected through/by the polarizer.  Assume normal operation is V-band in X-mode and Q-band in O-mode, so that one desires $\theta_p+\pi/2=\theta_{l,v}$ and $\theta_p=\theta_{l,q}$.

From the geometry we have that
\begin{equation}
\theta_{l,q}=\pi/2+ \tan^{-1}  \left(1/\sqrt{2} \tan \theta_{rot}  \right).
\end{equation}

\begin{figure}[!htbp]
\includegraphics[width=8.5 cm]{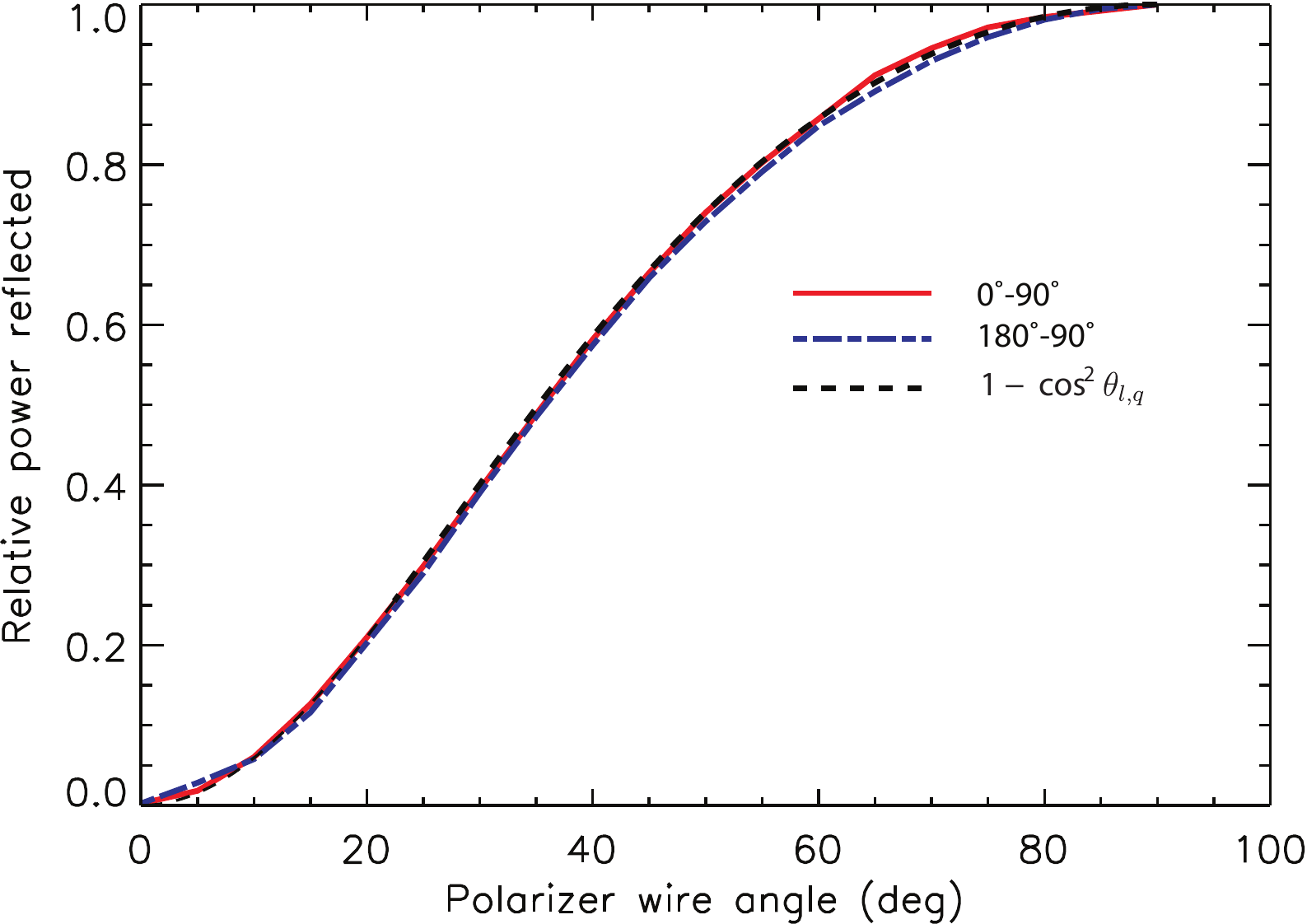}
\caption{\label{fig:polarizer}  Laboratory measurements of polarizer reflected power dependence on wire angle, with comparison to expectation.  Measurements for rotation greater than $90^{\circ}$ reversed to overlay data.}
\end{figure}

For an originally vertically polarized wave, the transmitted power should go as $\cos^2 \theta_{l,q}$.  Half the power will be transmitted at $\theta_{l,q}=45^{\circ}$, which requires from above that $\theta_{rot}=54.7^{\circ}$, or referenced to horizontal, $35.3^{\circ}$.  Comparison of expected reflected power to laboratory measurements shown in Fig.~\ref{fig:polarizer}, where angles beyond $90^{\circ}$ are reversed to overlay the data.  The laboratory measurements showed a high degree of polarization isolation and agreement with the predicted dependency.  This confirmed that even with its large size and bespoke construction, the polarizer worked as expected.  During experiments the polarization angle was adjusted to match planned plasma conditions.  In some cases with large differences, $\sim 10^{\circ}$ or more, two peaks could be observed in the signal, consistent with launch and detection of both polarizations simultaneously in these cases (although other mechanisms can also result in two measured peaks, even when polarization is well-matched).

\section{Data analysis \label{sec:analysis}}

The digitized data from the DBS systems were the output from quadrature mixers, where there is an in-phase, $I=A \cos \varphi$, and quadrature, $Q=A \sin \varphi$, component.  The amplitude, $A$, is the amplitude of the scattered electric field and the phase, $\varphi$, is the phase of the detected electric field referenced to a local oscillator.  Analyzed below is the complex electric field, $E=I+iQ$.  The two primary quantities of interest to extract from DBS data are the amplitude and Doppler shift of the localized signal coming from near the cutoff.  When there is little contribution to the total power from the near zero frequency component of the spectrum, thought to arise from scattering along the beam path (discussed further below), this can be accomplished easily via moments and integration of spectra.  For less ideal circumstances, fitting routines are used.  A three step algorithm is used to determine the amplitude and Doppler shift as reported in sections below.  A time series of sliding FFTs, using Hanning windows, are generated from the data.  Typically $2^{11}-2^{13}$ points are used for each spectrum.  First, moments of each spectrum are calculated to generate initial guesses for a fitting routine.  Second, similar to the procedure described in Ref.~\onlinecite{conway_interaction_2010}, the symmetric component of the spectrum is removed and an anti-symmetric double Gaussian is fit to $f(\omega)=E(\omega)-E(-\omega)$.  The near zero frequency component is usually close to symmetric and this procedure mostly removes it, as well as the background noise level.  This fitting procedure usually generates a good estimate of the Doppler shift, but in some cases the spectral shape is not very Gaussian (presumably due to refractive effects on the beam shape or due to poor localization in some cases) so the fit yields a poor estimate of the signal amplitude.  Therefore the third step is to use the Doppler shift from the fit to define a frequency window (so as to exclude the near zero frequency component), typically $\pm 1$ MHz around the peak fit.  The Doppler shift is then determined by the first moment of the bounded spectrum and the amplitude by its integration.  Consistency checks are performed and good quality data is taken to be when steps two and three generate similar values.  Error bars plotted below are calculated from the standard deviation of the values determined by the described procedure over a time period, typically between 1 and 5 ms.  Ray tracing is used as described in Sec.~\ref{sec:rays} to determined the scattering location, wavenumber, and alignment.  Whenever possible,  measurements from a Motion Stark Effect (MSE) diagnostic~\cite{conway_mast_2010} are used to constrain the EFIT equilibrium reconstructions.

\section{Validation of design methodology and cross-diagnostic comparison of velocity measurements\label{sec:validate}}

In this section, we assess the design calculations with comparisons to experimental measurements and present cross-diagnostic comparisons of velocity measurements, which ended up yielding unexpected results on core poloidal rotation.  One issue to note is that most of the pre-design calculations and inter-shot analysis during experiments was performed using magnetic equilibrium reconstructions dependent on magnetics data only, with monotonic safety factor profiles.  MAST plasmas often possess an elevated safety factor on axis and a region of reversed magnetic shear.  Particularly for accurate interpretation of core measurements, well-constrained equilibria are necessary.  Due to the inconsistent quality of the equilibrium reconstruction (MSE data is sometimes not available) and also large variations ($\sim 20$ cm) in the vertical position of the magnetic axis in different plasma scenarios, we report the launch angles at the steering mirror rather than effective incidence angles with the plasma.

\subsection{Investigation of toroidal angle scans}
Several data sets were acquired in repeated plasma conditions where the toroidal launch angle of the DBS mirror was systematically scanned at constant poloidal angle.  As shown in Fig.~\ref{fig:theta_path}, this should be expected to have a large impact on the mismatch angle, with only a small effect on the scattering wavenumber and location.  Investigation of these data sets allows the expectations set out in Sec.~\ref{sec:alignment} to be assessed.  In particular we expect that for a mirror setting aligned for the current flat top, there should be no measured signal early in the shot and the scattered signal should come into alignment as the current profile evolves.  There should also be a toroidal launch angle dependence of scattered power.

\begin{figure}[!htbp]
\includegraphics[width=6.0 cm]{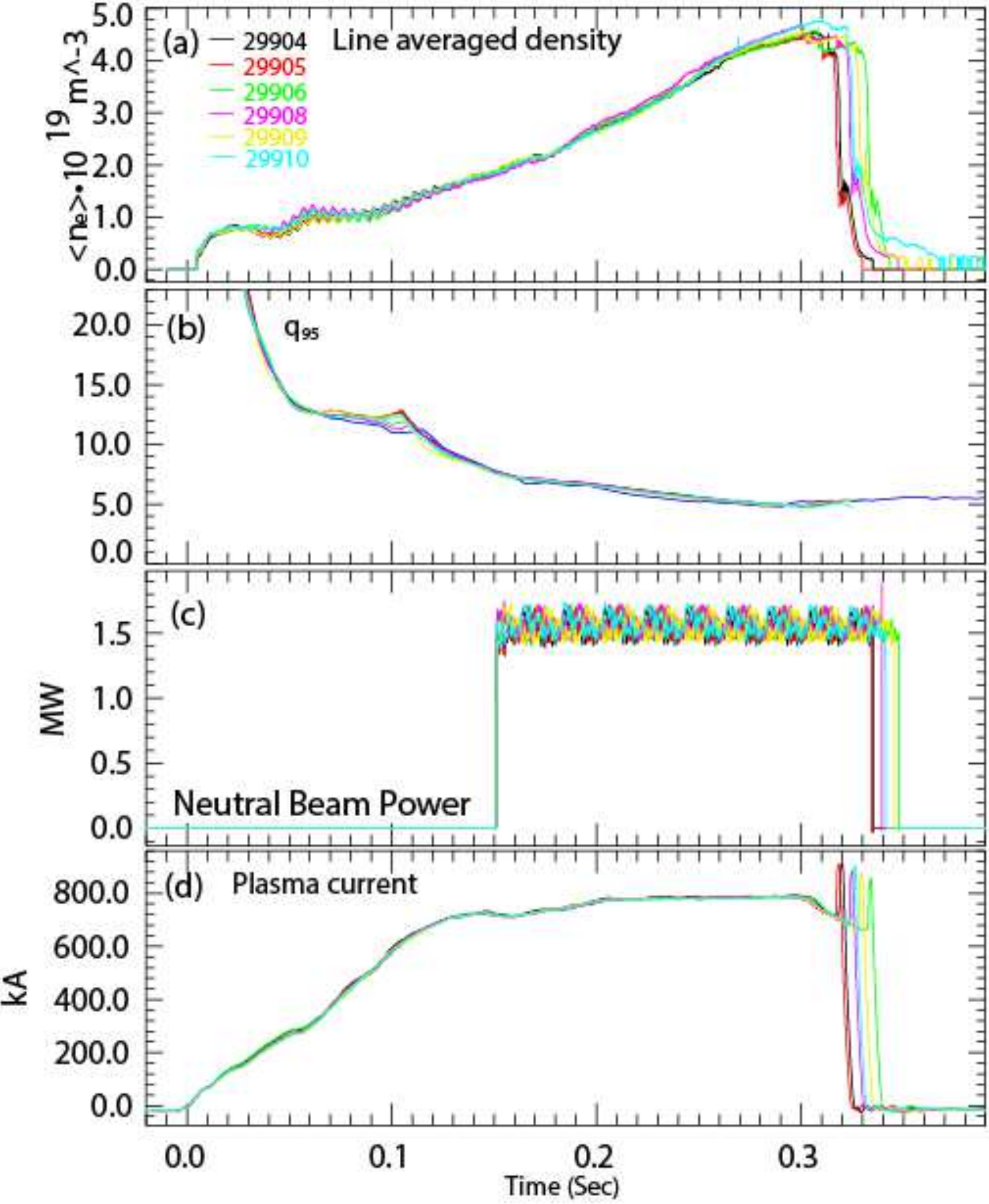}
\caption{\label{fig:traces1} Time history of equilibrium parameters for a sequence of MAST shots: (a) line averaged density, (b) edge safety factor, (c) injected neutral beam power, and (d) plasma current.}
\end{figure}

Figure~\ref{fig:traces1} shows the time history of equilibrium parameters for a sequence of MAST shots, where plasma conditions were held constant for diagnostic scans.  The plasmas were in L-mode.  During this sequence of shots the DBS poloidal launch angle was held constant at $-4^{\circ}$ from horizontal and the toroidal angle was scanned in $1^{\circ}$ increments from $-1^{\circ}$ to $-6^{\circ}$ about the mirror axis.  The effective toroidal angle referenced to the plasma is discussed below.  Good quality MSE data was only acquired for a subset of the shots; however, the difference between the magnetics-only equilibrium reconstruction and the MSE-constrained reconstruction was larger than the shot-to-shot variation at a particular time.  Therefore we use the shot with the lowest uncertainty MSE data for ray tracing calculations below.

\begin{figure}[!htbp]
\includegraphics[width=8.5 cm]{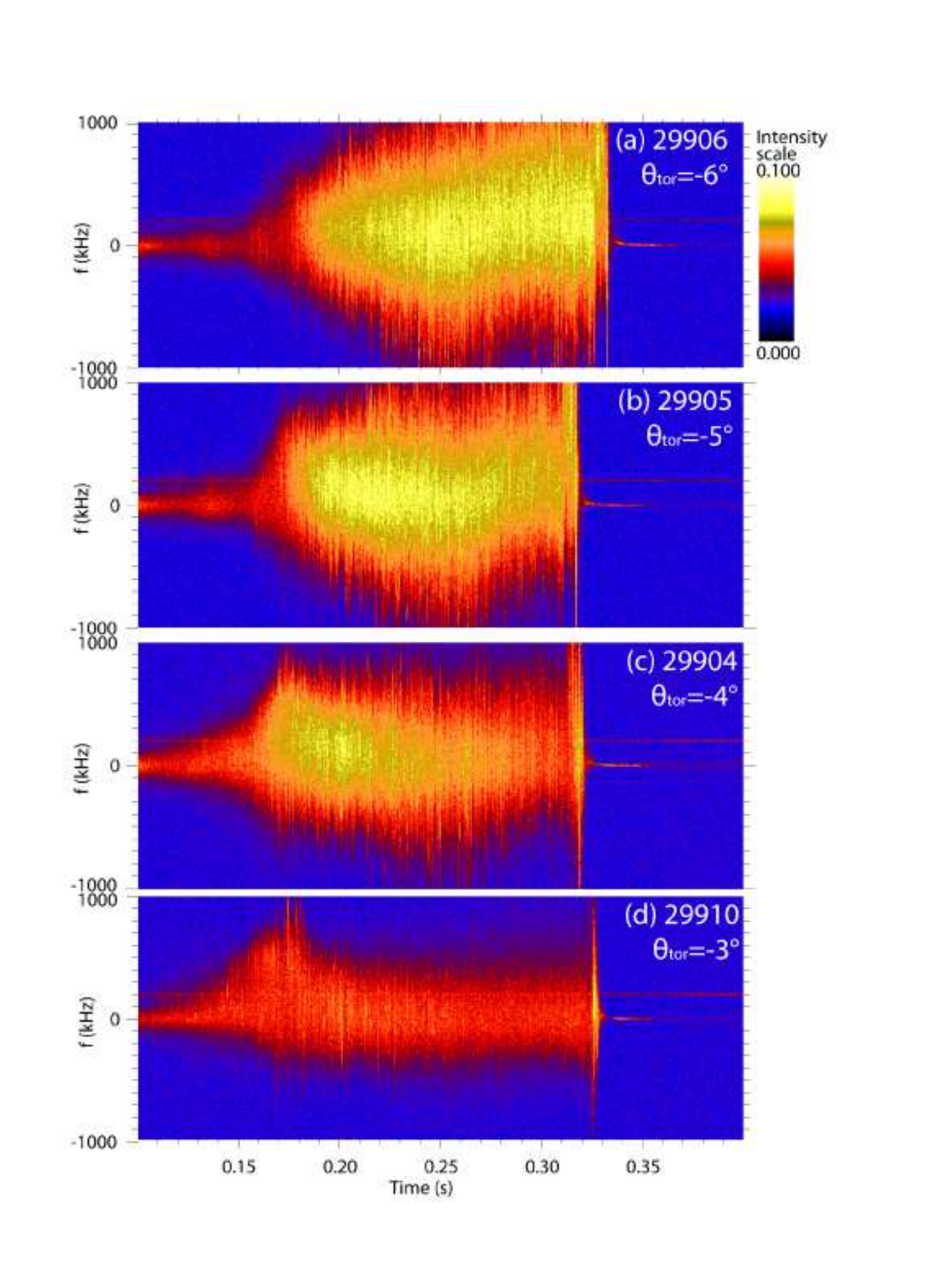}
\caption{\label{fig:scan1} Spectrograms of 47.5 GHz, X-mode DBS channel in sequence of repeated shots.  Plots are of detected complex electric field in logarithmic scale.  All cases were launched at $-4^{\circ}$ poloidal angle.  Toroidal launch angles about mirror axis were (a) $-6^{\circ}$, (b) $-5^{\circ}$, (c) $-4^{\circ}$, and (d) $-3^{\circ}$.  Same scale used for all plots.}
\end{figure}

Figure~\ref{fig:scan1} shows the scattered electric field from the 47.5 GHz channel, which was oriented for X-mode polarization, in four of the shots.  The impact of the scattering misalignment has a large effect on the measurement.  The scattering location was $\sqrt{\psi} \approx 0.90$ and the scattering wavenumber was $k_{n, \bot} \approx 7\ \mathrm{cm}^{-1}$, with the toroidal angle scan changing the scattering wavenumber by order $10\%$.  The qualitative changes from panel-to-panel can be qualitatively interpreted with the aid of the ray tracing results in Fig.~\ref{fig:theta_time}.  In all cases, the Doppler shifted signal does not appear until the after 150 ms, which is consistent with Fig.~\ref{fig:theta_time}(a), where the mismatch angle reduces to close to zero as the scattering comes into alignment.  The effect of changing the toroidal angle on the time history of the mismatch angle is essentially to vertically shift a plot such as Fig.~\ref{fig:theta_time}(a).  This can be seen in the DBS data in Fig.~\ref{fig:scan1}, where from (a) to (d) the alignment condition is met earlier as the toroidal angle is scanned.  Fig.~\ref{fig:scan1}(a) is similar to Fig.~\ref{fig:theta_time}(a), where the mismatch angle is well-matched to the asymptotic pitch angle as the current profile approaches steady-state.  In Figs.~\ref{fig:scan1}(b-d), most clearly in (c), the mismatch angle passes through zero and the scattering comes into optimal alignment then goes out of alignment.  All shots have a contribution to the signal around zero frequency, which is thought to be from non-localized high-k scattering along the entire path, which is always present to some degree.  We conclude that it is most likely high-k backscattering since it is only present when there is a plasma and it is always observed when there is a plasma, regardless of the scattering alignment near the cutoff. The low frequency of the peak is consistent with high-k backscattering along the beam path, which would mostly be $k_r$ and so not Doppler shifted. 

\begin{figure}[!htbp]
\includegraphics[width=8.5 cm]{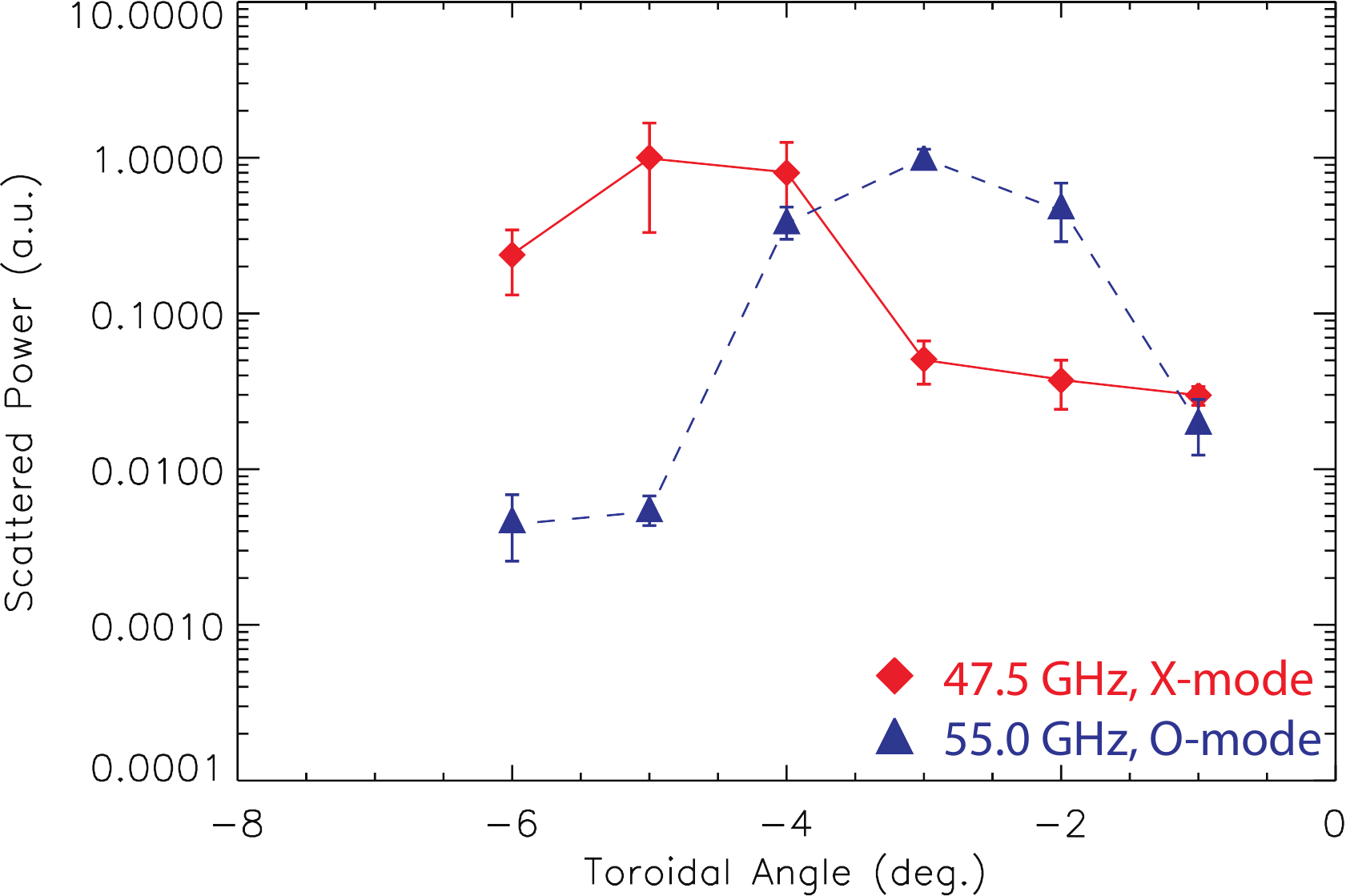}
\caption{\label{fig:flucts1} Scattered power at 190 ms in shots $29904-29906,\ 29908-29910$ from two DBS channels versus toroidal launch angle.  Red diamonds are 47.5 GHz, X-mode at $\sqrt{\psi} \approx 0.90$ and $k_n \approx 7\ \mathrm{cm}^{-1}$.  Blue triangles are 55.0 GHz, O-mode at $\sqrt{\psi} \approx 0.70$ and $k_n \approx 9\ \mathrm{cm}^{-1}$.  Data from each channel is separately normalized to the maximum from the scan.}
\end{figure}

We now compare the measured scattered power at the same time in different shots.  Figure~\ref{fig:flucts1} shows the dependence of received scattered power on toroidal launch angle for two of the DBS channels at 190 ms, averaged over 5 ms.  Red diamonds are 47.5 GHz, X-mode at $\sqrt{\psi} \approx 0.90$ and $k_n \approx 7\ \mathrm{cm}^{-1}$.  Blue triangles are 55.0 GHz, O-mode at $\sqrt{\psi} \approx 0.70$ and $k_n \approx 9\ \mathrm{cm}^{-1}$.  The abscissa is the toroidal mirror angle about its rotation axis.  As expected, both channels show a clear maximum as toroidal angle is scanned, which levels out to detection levels at large mismatch angles (due to inability to distinguish from the near zero frequency contribution in this case, not from system noise levels).  For both channels, there is a large drop in detected power for $2^{\circ}$ or more from the maximum, which is consistent with estimates from pre-installation design calculations in Sec.~\ref{sec:design}.  It is also notable that the maximum detected power occurs at different toroidal angles for the two channels, which is consistent with Fig.~\ref{fig:theta_freq} and is due to radial variation of the magnetic field pitch angle.  The higher frequency channel, for which the cutoff is at a smaller radii, is best matched for a smaller toroidal angle as the magnetic field pitch angle is smaller closer to the magnetic axis.  This result also illustrates one of the challenges for implementing DBS in a spherical tokamak, where although there is overlap in toroidal angles over which both channels have high signal levels, both cannot be maximized at the same time.

It is clear from Fig.~\ref{fig:flucts1} that the toroidal alignment is an important effect.  To compare fluctuation levels or construct wavenumber spectra in a spherical tokamak in different plasma conditions, which in general can have different mismatch angles, it must be taken into account.  The effect is less of a concern for standard, large aspect ratio tokamaks, owing to their smaller variation of magnetic field pitch angle.  From the framework in Sec.~\ref{sec:alignment}, three important quantities must be known for the misalignment to be corrected: the size of the beam, the scattering wavenumber, and the mismatch angle.  The wavenumber can be determined by ray tracing.  The size of the beam can be estimated using beam tracing or fullwave calculations, estimated from vacuum measurements of beam profiles, or may in fact be determined from the data such as in Fig.~\ref{fig:flucts1}, since the beam size should impact the angular range of the alignment.  The mismatch angle can also be determined from ray tracing, but can be highly sensitive to the equilibrium reconstruction.  Furthermore, it was uncovered that the toroidal angle of the mirror launch was slightly different than expected from mechanical drawings, which was revealed when attempting to compare measurements at positive and negative poloidal angles (different toroidal angles were necessary at the same magnitude of poloidal angle to produce similar data in balanced up-down symmetric plasmas).  The mechanical placement of the support frame and quasi-optical components relative to nearby reference objects were confirmed to within better accuracy than could account for the offset.  Vacuum tests helped little to resolve the issue, but were performed with toroidal field coils and Ohmic windings inactive; those corresponding mechanical stresses might account for part of the offset.  In steep density gradients non-WKB effects might give rise to an apparent toroidal misalignment~\cite{holzhauer_note_2006}, which might be relevant to the H-mode pedestal.  The effect was always present and did not appear to vary with plasma conditions.  No single cause was identified for the offset, but it is possible several small effects are being compounded.  

\begin{figure}[!htbp]
\includegraphics[width=8.5 cm]{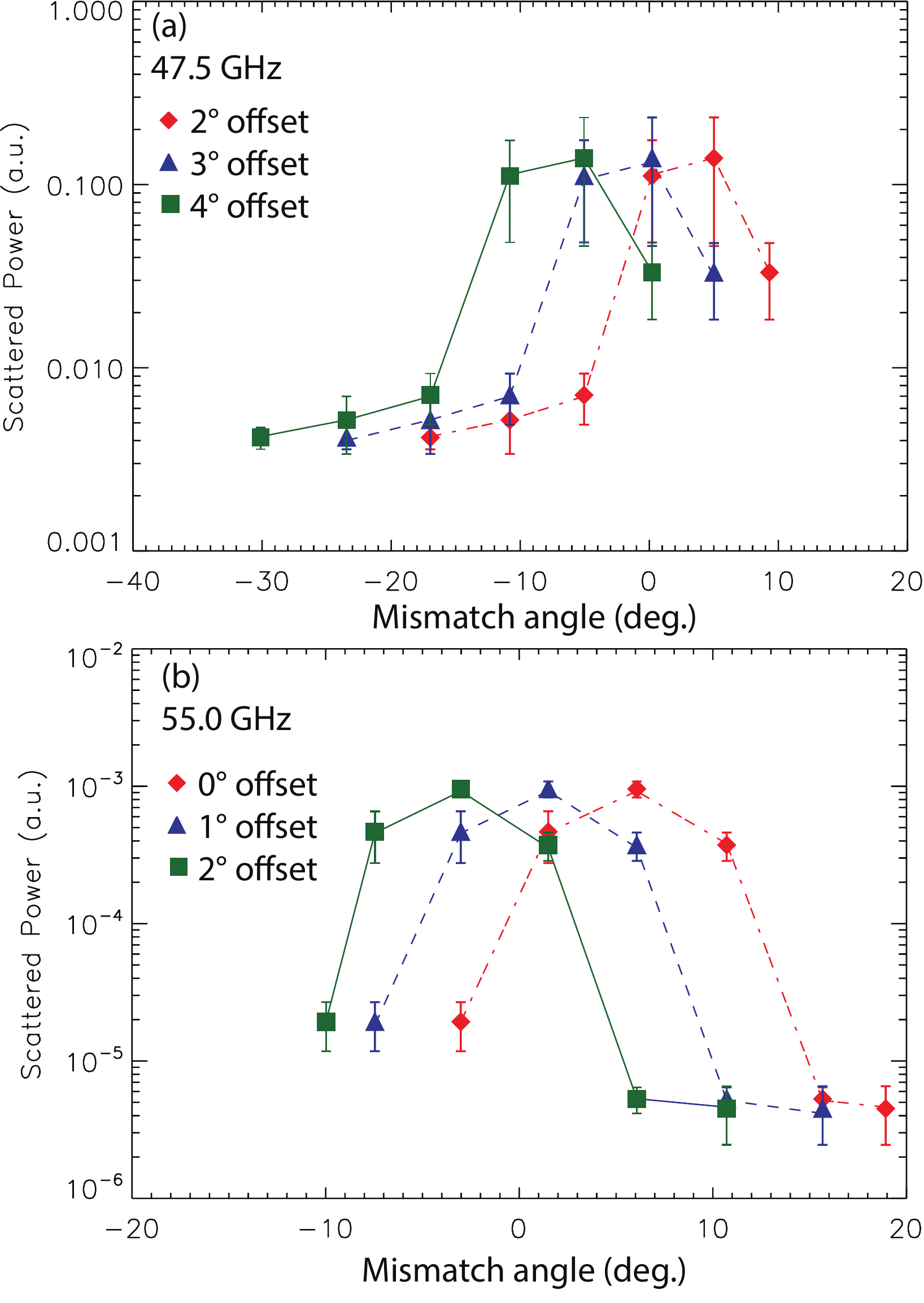}
\caption{\label{fig:tor_offset} Scattered power at 190 ms in shots $29904-29906,\ 29908-29910$ from two DBS channels, (a) 47.5 GHZ and (b) 55.0 GHz versus scattering mismatch angle from ray tracing, with the effect of a toroidal angle offset on the calculation shown for each case.}
\end{figure}

The offset issue is illustrated by Figure~\ref{fig:tor_offset}, which plots the data from Fig.~\ref{fig:flucts1} against ray tracing calculations of the scattering mismatch angle.  The scattered power should peak at a mismatch angle of zero degrees.  An offset of $\sim 2-3^{\circ}$ would be needed for the 47.5 GHz channel and $\sim 1^{\circ}-2^{\circ}$ for the 55.0 GHz channel.  A mechanical misalignment should affect all channels with the same offset.  Note again that calculation of the mismatch angle relies on ray tracing, which uses equilibrium reconstruction from EFIT.  The calculations in Fig.~\ref{fig:tor_offset} use MSE-constrained reconstructions. A $1^{\circ}$ difference in toroidal angle, with the MAST DBS geometry, makes about a $5^{\circ}$ difference in $\theta_{mis}$.  The difference between MSE-constrained and magnetics only EFIT in the core can be $\sim 10^{\circ}$ in $\theta_{mis}$.  The shot-to-shot variation in $\theta_{mis}$, for a particular frequency and time in the toroidal angle scan, is $\sim 2^{\circ}-3^{\circ}$.  To arrive at the point where the scattering mismatch can be reliably and robustly corrected for in its effect on the scattered power, and to estimate the uncertainty in that correction, will require examination of a data set beyond the scope of the work presented here.  This would be necessary for study of parametric dependencies of the wavenumber spectrum of turbulence, for instance.  Calculation of the scattering wavenumber and position depend weakly on the mismatch, so evaluation of the radial electric field is not strongly impacted by this issue.  We currently take the offset to be $2^{\circ}\pm 1^{\circ}$.

\subsection{Cross-diagnostic comparison of velocity measurements}
Cross-diagnostic comparisons are important to validate new measurements and analysis procedures.  Two additional independent measurements sensitive to plasma rotation are available at MAST.  The first is a charge exchange recombination spectroscopy (CXRS) diagnostic~\cite{conway_high-throughput_2006}, which provides toroidal rotation velocity, ion temperature, and ion density from carbon impurity measurements.  Poloidal rotation in MAST is usually small~\cite{field_comparison_2009} (and also close to or within measurement uncertainties), and is omitted from the analysis here.  The second is a beam emission spectroscopy (BES) diagnostic~\cite{field_beam_2012}, which can measure the motion of ion-scale turbulence, the velocity of which can be determined through time delay estimation~\cite{ghim_measurement_2012}.

Force balance requires that for each species in the plasma
\begin{equation} \label{eqn:er}
E_r=\frac{\nabla P_i}{e Z_i n_i} + v_{\phi,i}B_{\theta}-v_{\theta,i} B_{\phi},
\end{equation}
where $E_r$ is the radial electric field, $P_i$ is the pressure of species $i$ and $n_i$ is its density.  The electric charge is $e$, $Z_i$ is the atomic charge number, and for the velocity and magnetic field components, $\phi$ is the toroidal direction and $\theta$ is the poloidal direction.  Doppler backscattering measures a Doppler shift induced by scattering from turbulent structures in the plasma, which has contributions from both the plasma $E \times B$ drift velocity and the phase velocity of the turbulence, $v_{turb}=v_{E \times B}+v_{phase}$.  However, $v_{phase}$ would be expected to be on the order of the diamagnetic velocity, $v_{dia,s}=\nabla P_s/q B n_s$ for species $s$, but could in principle be in either the ion or electron direction.  For most conditions the $E \times B$ velocity dominates and $v_{turb}\approx v_{E \times B}$.  This then provides a measurement of the radial electric field from the expression $\mathbf{v}_{E \times B}=\bm{E} \times \bm{B}/B^2$.  Therefore if the ion pressure gradient and poloidal rotation velocity are both small, $v_{turb}B/B_{\theta} \approx v_{\phi}$, and DBS and CXRS toroidal velocity measurements can be compared. Similar to DBS, BES measures the apparent velocity of the turbulence.  Due to viewing the turbulence in a poloidal plane, an additional geometric factor enters~\cite{ghim_measurement_2012} and when the toroidal rotation dominates, $v_{turb,BES}B_{\phi}/B_{\theta} \approx v_{turb,DBS}B/B_{\theta} \approx v_{\phi,CXRS}$.

\begin{figure}[!htbp]
\includegraphics[width=8.5 cm]{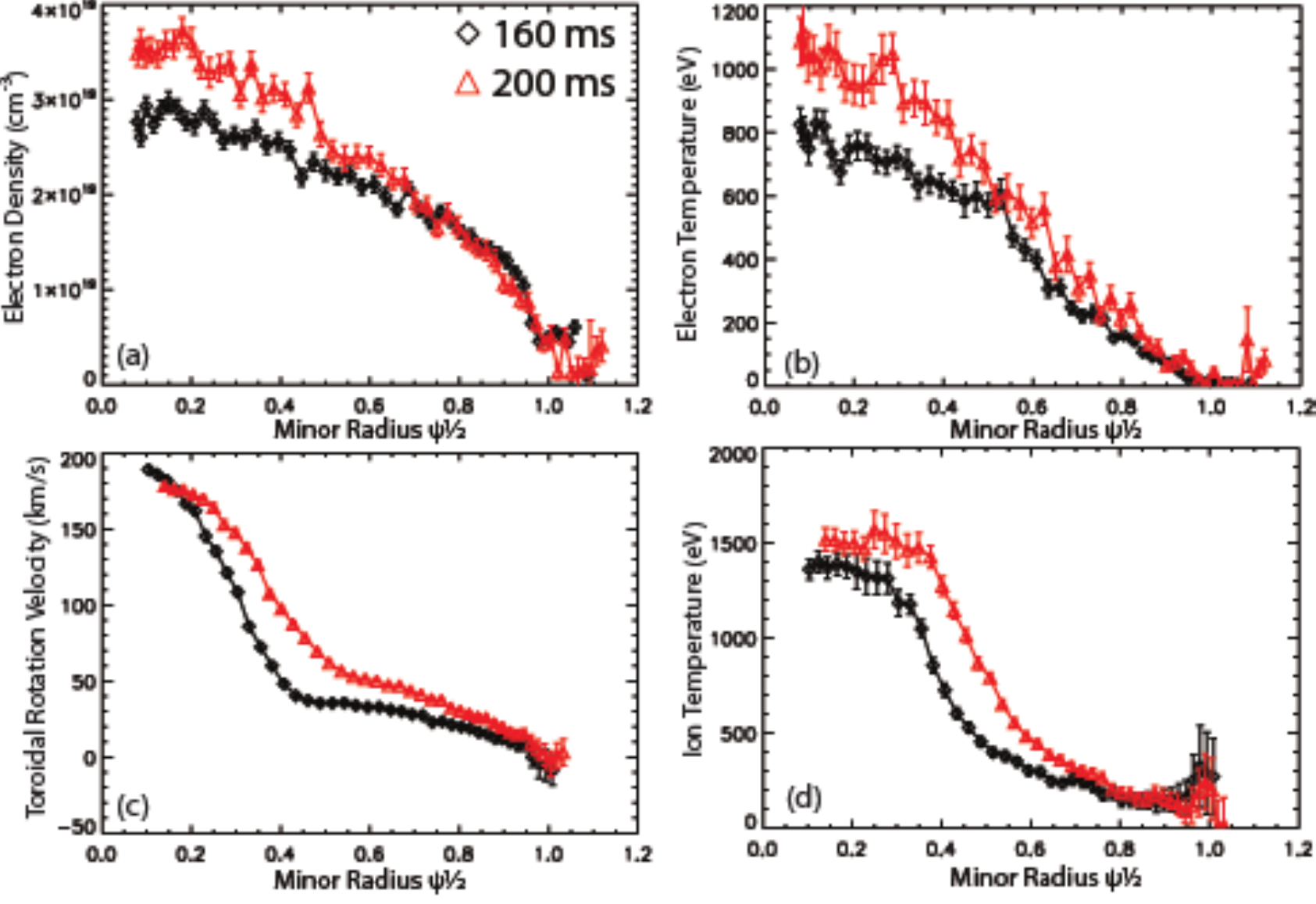}
\caption{\label{fig:profiles30113} Equilibrium profile data from shot 30113 and 160 ms and 200 ms: (a) electron density, (b) electron Temperature, (c) toroidal rotation velocity, and (d) ion temperature mapped to the outboard midplane and plotted against the square root of the normalized poloidal flux.}
\end{figure}

For the comparison, we consider two times from an L-mode MAST shot with an internal transport barrier.  ITBs have been previously studied in MAST~\cite{field_plasma_2011}.  Examining an L-mode discharge with an ITB is useful for this comparison since the L-mode density profile allows DBS to probe a wide radial region and the large ion temperature gradient in the ITB should have an impact on the radial electric field through the pressure gradient term in Eq.~\ref{eqn:er}.  Figure~\ref{fig:profiles30113} shows the equilibrium electron density and temperature from Thomson scattering, and the toroidal rotation velocity and ion temperature from CXRS.  MSE-constrained EFITs are used for the mapping to flux coordinates for the ray tracing wavenumbers used to infer velocities below.  The profiles in Fig.~\ref{fig:profiles30113} show the ITB, which results in a large ion temperature gradient in the region $\sqrt{\psi} \approx 0.4-0.6$, that evolves between the two times under consideration.

\begin{figure}[!htbp]
\includegraphics[width=8.5 cm]{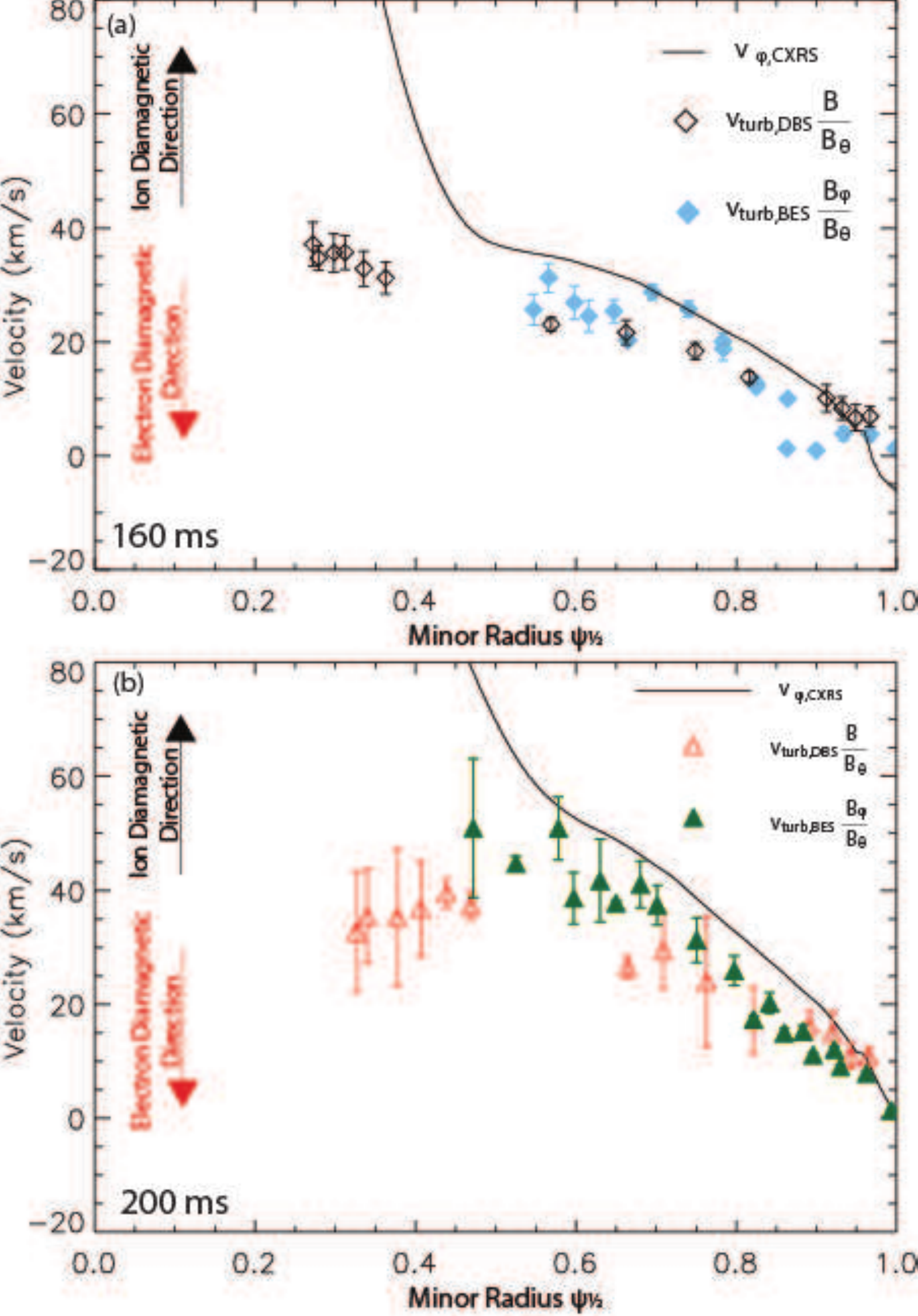}
\caption{\label{fig:3diag} Three way comparison between DBS, BES, and CXRS toroidal rotation velocity from MAST shot 30113 (BES also uses 30114 and 30015) at (a) 160 ms and (b) 200 ms.}
\end{figure}

Figure~\ref{fig:3diag} shows the three way comparison between DBS, BES, and CXRS toroidal rotation velocity.  The DBS and CXRS data are from 30113.  The shot was repeated twice and the BES viewing location was moved; the BES data is from 30113, 30114, and 30115.  As noted above, if toroidal rotation dominates compared to the ion pressure gradient, poloidal rotation, and the turbulence phase velocities (since BES and DBS measure different spatial scales, the phase velocities would, in general, be different), then $v_{turb,BES}B_{\phi}/B_{\theta} \approx v_{turb,DBS}B/B_{\theta} \approx v_{\phi,CXRS}$, and the three measurements should give the same result.  What happens in Fig.~\ref{fig:3diag} is that all three give similar results for $\sqrt{\psi} > 0.5$, with some interesting small differences.  All three also show an increase across the minor radius from 160 ms to 200 ms.  In the plots, the electron diamagnetic velocity direction (hereafter shortened to ``electron direction'') is towards the negative direction and the ion diamagnetic velocity direction (``ion direction'') is in the positive direction.  The sign convention is such that co-current (also the direction of neutral beam injection) toroidal rotation contributes towards the positive direction, and the ion pressure gradient term contributes towards the negative direction.  Both the DBS and BES measurements generally fall below the CXRS, which would be consistent with a small contribution from the ion pressure gradient term.  Near the edge ($\sqrt{\psi} \gtrsim 0.85$) the DBS measurements are generally slightly shifted in the ion direction direction than BES, while in the core ($\sqrt{\psi} \lesssim 0.85$) the DBS is slightly shifted in the electron direction.  The latter is what one would generically expect, since BES measures ion-scale turbulence ($k_{\bot} \rho_i \lesssim 1$) most commonly thought to be driven by ion temperature gradient modes that typically propagate in the ion diamagnetic direction, while DBS measures smaller scale fluctuations possibly related to trapped electron modes or electron temperature gradient modes, both of which propagate in the electron direction.  Speculatively, the observation that the opposite of this expectation is true near the edge might indicate a difference in the turbulence drive at that location.  Any specific conclusions will require comparison to modeling and is deferred to future work; however, the point here is that differences between the turbulence velocities would be expected, so it is not a surprise that $v_{turb,BES}B_{\phi}/B_{\theta}$ and $ v_{turb,DBS}B/B_{\theta}$ are slightly different, and the difference is on the order of the diamagnetic velocity (\textit{i.e.} order of contribution of the pressure gradient term). The scattering wavenumber for DBS ranges from about $4\ \mathrm{cm}^{-1}$ for the outermost point to about $22\ \mathrm{cm}^{-1}$ for the innermost, even at the outermost point this is still about a factor of two larger wavenumber than BES. For measurement locations inside the ITB, a discrepancy starts to appear between CXRS and the turbulence measurements from DBS and BES.

\begin{figure}[!htbp]
\includegraphics[width=8.5 cm]{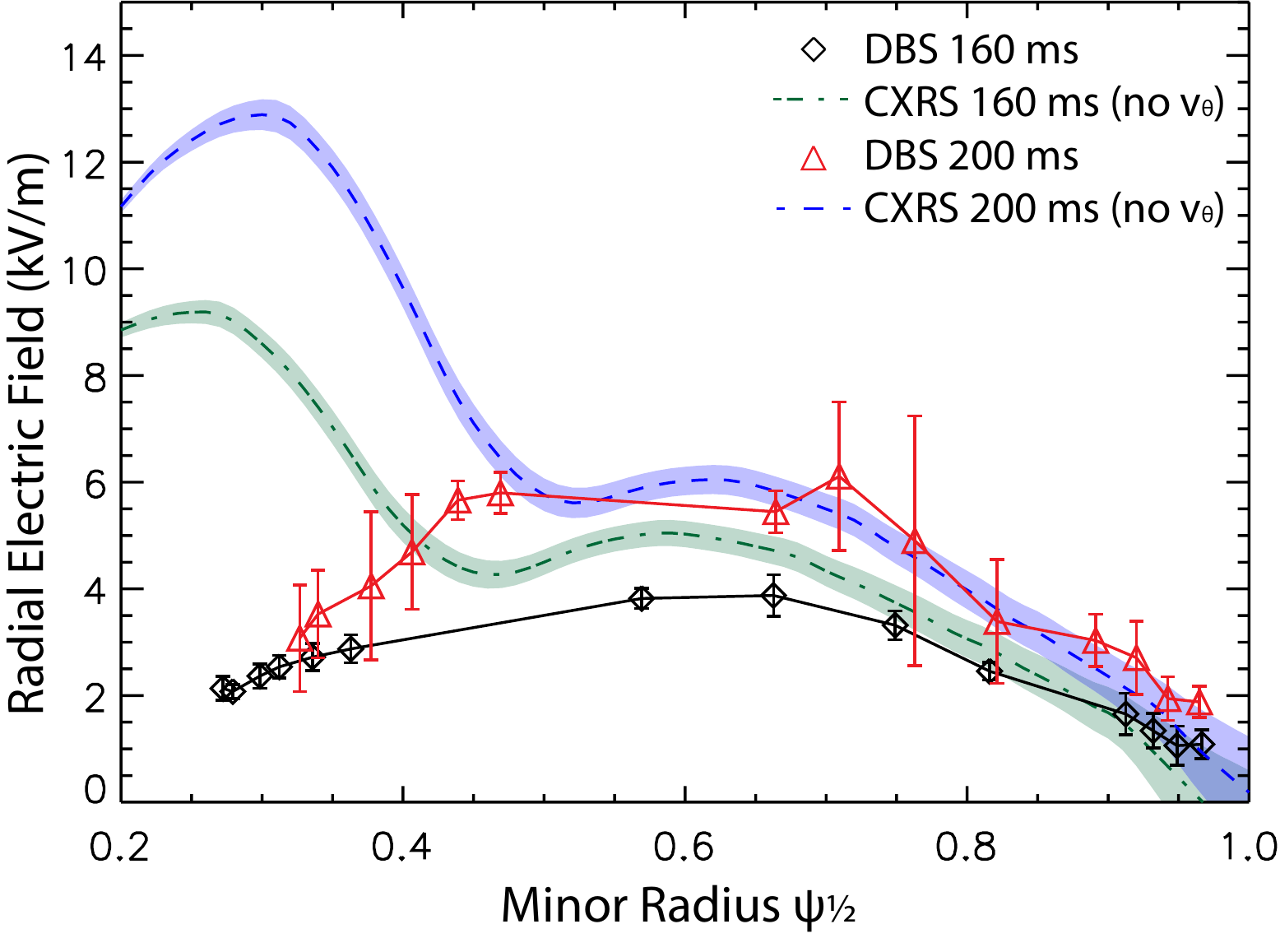}
\caption{\label{fig:compare_er} Comparison of inferred radial electric field from DBS and CXRS (excluding poloidal rotation), in MAST shot 30113 at 160 ms and 200 ms.  Shaded regions indicate uncertainty bounds.}
\end{figure}

To investigate the observed differences in Fig.~\ref{fig:3diag}, Fig.~\ref{fig:compare_er} compares the inferred radial electric field from DBS (assuming the turbulence phase velocity is small) and from CXRS (including the ion pressure gradient term, but excluding poloidal rotation, which was not available).  The ion pressure gradient term contributes up to -2 kV/m in the ITB.  The shaded region around CXRS indicates the uncertainty range.  The result shows that for most positions $\sqrt{\psi} > 0.4$ the two measurements are within uncertainties.  All points $\sqrt{\psi} > 0.4$ could be made to agree by assuming a poloidal velocity of 2 km/s or less, which is of the order expected for neoclassical poloidal rotation in MAST~\cite{field_comparison_2009}.  The points $\sqrt{\psi} < 0.4$, inside the ITB, are potentially more interesting.  The difference inside the ITB implies a poloidal velocity of up to $\sim 15$ km/s, which is larger than would be predicted by standard neoclassical poloidal rotation calculations, but observations of similar magnitude have been made before in MAST (cf. Fig. 8 of Ref.~\onlinecite{field_comparison_2009}).  Large poloidal rotation connected with ITBs was previously observed in JET~\cite{crombe_poloidal_2005} as well.  Measurements in DIII-D have also found poloidal rotation exceeding neoclassical predictions in low collisionality plasmas~\cite{grierson_collisionality_2013}.  Alternatively, there could be a very large phase velocity for the high-k DBS measurements; however, we consider this to be unlikely as the largest disagreements are not actually observed at the location of the largest pressure gradient and agreement is observed in some cases with similarly large wavenumbers.  The locations where a large poloidal velocity is inferred are approaching the magnetic axis, where finite orbit width and potato orbit effects might be significant, which were not included in previous comparisons~\cite{field_comparison_2009}.

\section{High-k DBS measurements} \label{sec:highk}

\begin{figure}[!htbp]
\includegraphics[width=6.0 cm]{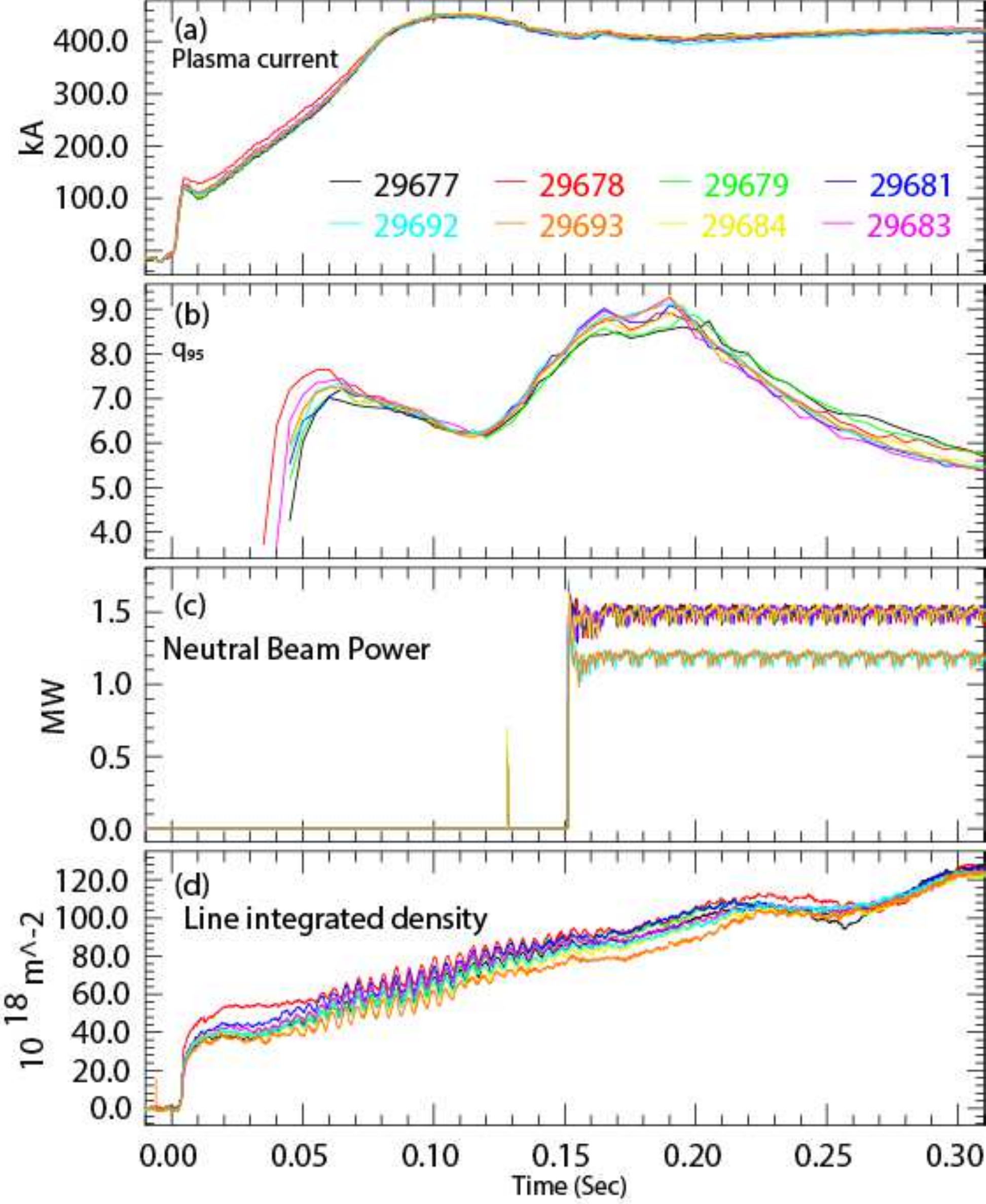}
\caption{\label{fig:traces2} Time history of equilibrium parameters for a sequence of MAST shots: (a) plasma current, (b) edge safety factor, (c) inject neutral beam power, and (d) line integrated density.}
\end{figure}

\begin{figure}[!htbp]
\includegraphics[width=6.0 cm]{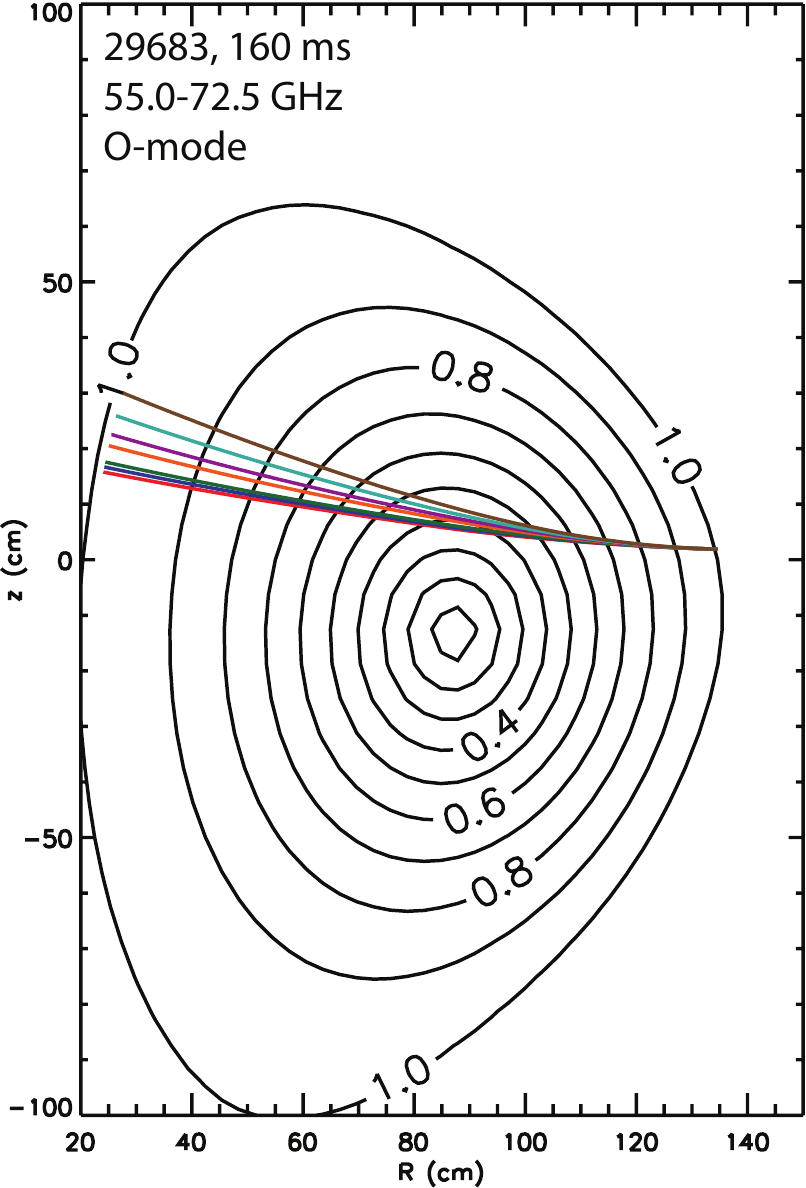}
\caption{\label{fig:kighk_rays} Ray tracing for 55.0 to 72.5 GHz O-mode channels showing trajectory for high-k measurements.  Contours of square root of the poloidal flux are plotted. Poloidal launch angle $1^{\circ}$ and toroidal launch angle $2^{\circ}$}
\end{figure}

A second toroidal angle scan was performed in a plasma where V-band, O-mode channels were sensitive to high-k fluctuations.  The shot-to-shot consistency of equilibrium parameters is displayed in Fig.~\ref{fig:traces2}.  Different resonant magnetic perturbation coil combinations were applied from about 230 ms onward, but prior to that all shots had similar L-mode conditions, with some small variation in NBI power.  In O-mode, the V-band frequencies are actually above cutoff and the beam trajectory is only slightly deflected, as can be seen in the ray tracing in Fig.~\ref{fig:kighk_rays}.  These shots had no good MSE data, so detailed assessment of the mismatch angle is not performed here.  The minimum ray perpendicular index of refraction for these cases, which occurs at $\sqrt{\psi} \approx 0.4$, is only $k_{i}/k_0 \approx 0.6-0.8$, but toroidal steering enables measurement of a well-defined Doppler shifted peak.  The DBS mirror was set to $1^{\circ}$ poloidal launch and the plasmas were vertically shifted as in Fig.~\ref{fig:kighk_rays}.  The toroidal DBS mirror angle was scanned between $1^{\circ}$ and $6^{\circ}$.  Note that the sign of the Doppler shift has the opposite sense of Fig.~\ref{fig:scan1}; there the beam was launch down and to the right of normal incidence, here launch is up and to the left (as viewed from the mirror towards the plasma).

\begin{figure}[!htbp]
\includegraphics[width=8.5 cm]{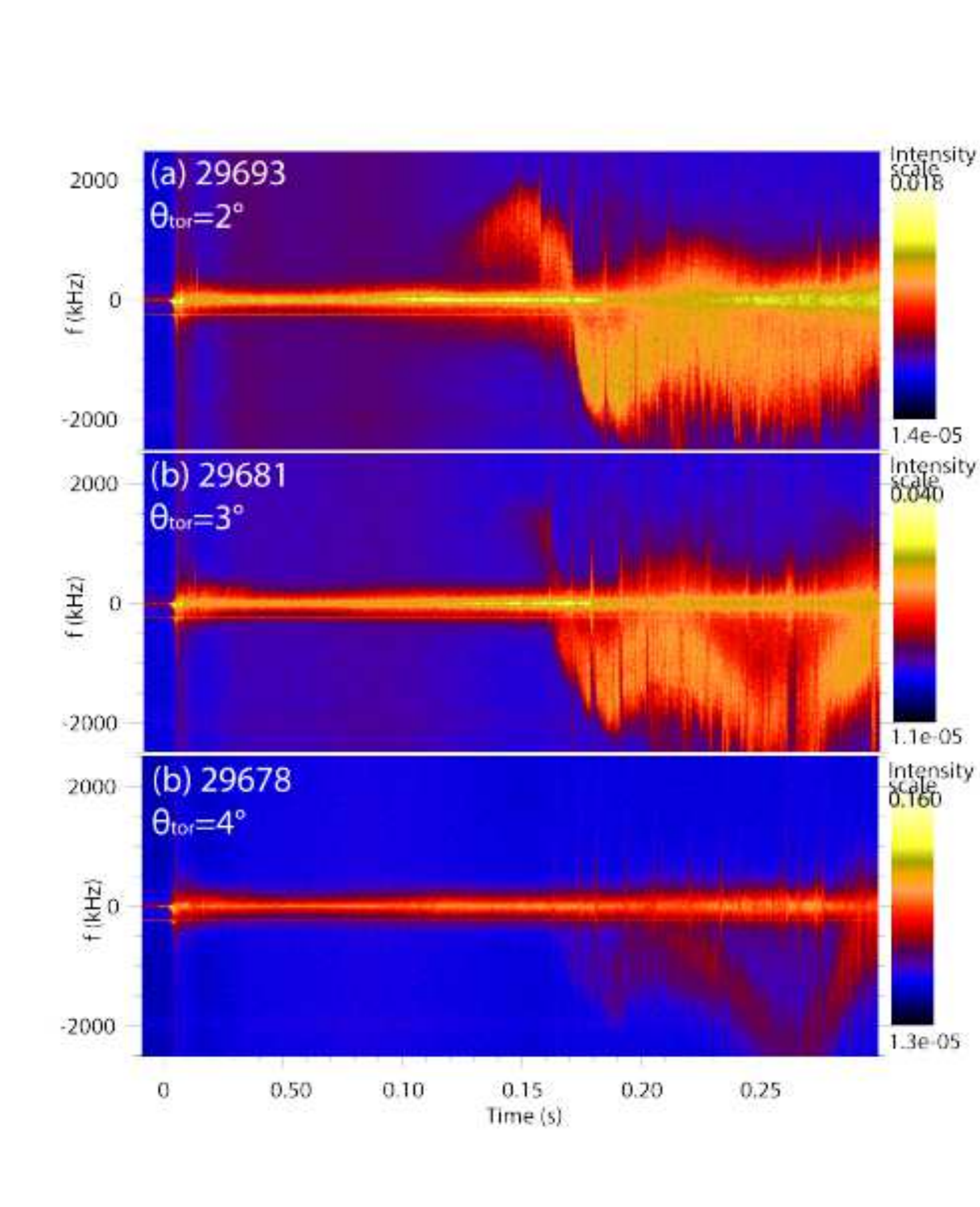}
\caption{\label{fig:scan2} Spectrograms of 55.0 GHz, O-mode DBS channel in sequence of shots with repeated conditions until about 230 ms.  Plots are of detected complex electric field in logarithmic scale.  All cases were launched at $1^{\circ}$ poloidal angle.  Toroidal launch angles about mirror axis were (a) $2^{\circ}$, (b) $3^{\circ}$, and (c) $4^{\circ}$.}
\end{figure}

Figure~\ref{fig:scan2} shows the effect of the toroidal angle scan for this case.  Again, similar to Fig.~\ref{fig:scan1}, the Doppler shifted peak is not measured until the scattering angle comes into alignment, which changes for different toroidal launch angles.  Also notable is that the sign of the measured flow changes after NBI is applied at 150 ms, where the rotation is initially in the counter-$I_p$ direction in the Ohmic phase, then changes to co-$I_p$ with the addition of momentum input from the neutral beams.  At 150 ms the scattering wavenumber is $k_{n,\bot} \approx 16 \ \mathrm{cm}^{-1}$ or $k_{n,\bot} \rho_i \approx 10$.  The highest frequency channel for which data was acquired was 72.5 GHz at $k_{n,\bot} \approx 24 \ \mathrm{cm}^{-1}$ or $k_{n,\bot} \rho_i \approx 15$, which is well above ion-scale fluctuations and into the electron scale.  Due to the grazing beam trajectories, all channels measure close to the same radial location.  These measurements demonstrate that with toroidal steering DBS is able to measure both intermediate-k and high-k density fluctuations.  There is a similar maximum in scattered power as a function of toroidal angle as shown in Fig.~\ref{fig:flucts1}; however, without MSE for theses shots we cannot at this time assess the mismatch angle with accuracy.  

\subsection{Localization of high-k measurements}

The high-k DBS measurements, with relatively small change in the beam index of refraction compared to normal DBS measurements, are only possible for general plasma conditions with the optimization available with two dimensional steering.  It should be noted that the self-alignment effect for O-mode propagation is also beneficial for these measurements.  This occurs due to the toroidal drift of a microwave beam in a plasma, where X-mode drifts away from being perpendicular to the magnetic field, while O-mode drifts towards being perpendicular~\cite{gourdain_assessment_2008,gourdain_application_2008}.  Despite these considerations, localization of the measurement should still be considered carefully; indeed, for toroidal angles such that there is very poor alignment it is possible to find pathological examples where there is no clear Doppler shifted peak, but there is instead a broad spectrum, smeared-out over several megahertz like one would expect from poor localization resulting in sampling a range of radii and velocities.

The high-k beam trajectories are typically similar to Fig.~\ref{fig:kighk_rays}, where there is only slight deflection of the beams.  These cases can even be thought of as similar to traditional scattering arrangements  (\textit{i.e.} $\omega >> \omega_{pe},~\omega_{ce}$ ), but with the localization benefit of DBS.  Interpretation of the received power as DBS data requires localization along the path, near to the radii of smallest perpendicular index of refraction along the path.  Since the highest V-band frequency is almost 50\% higher than the lowest, there is a significant difference in wavenumber even when there is little difference in the path.  Therefore if the measurement is localized, we expect that for cases like Fig.~\ref{fig:kighk_rays} each of the channels should have significantly different Doppler shifts and associated scattering wavenumbers, but in combination they should result in the same local velocity for the turbulence via  $\omega_{DBS}=k_{n,\bot}v_{turb}$.

\begin{figure}[!htbp]
\includegraphics[width=6.0 cm]{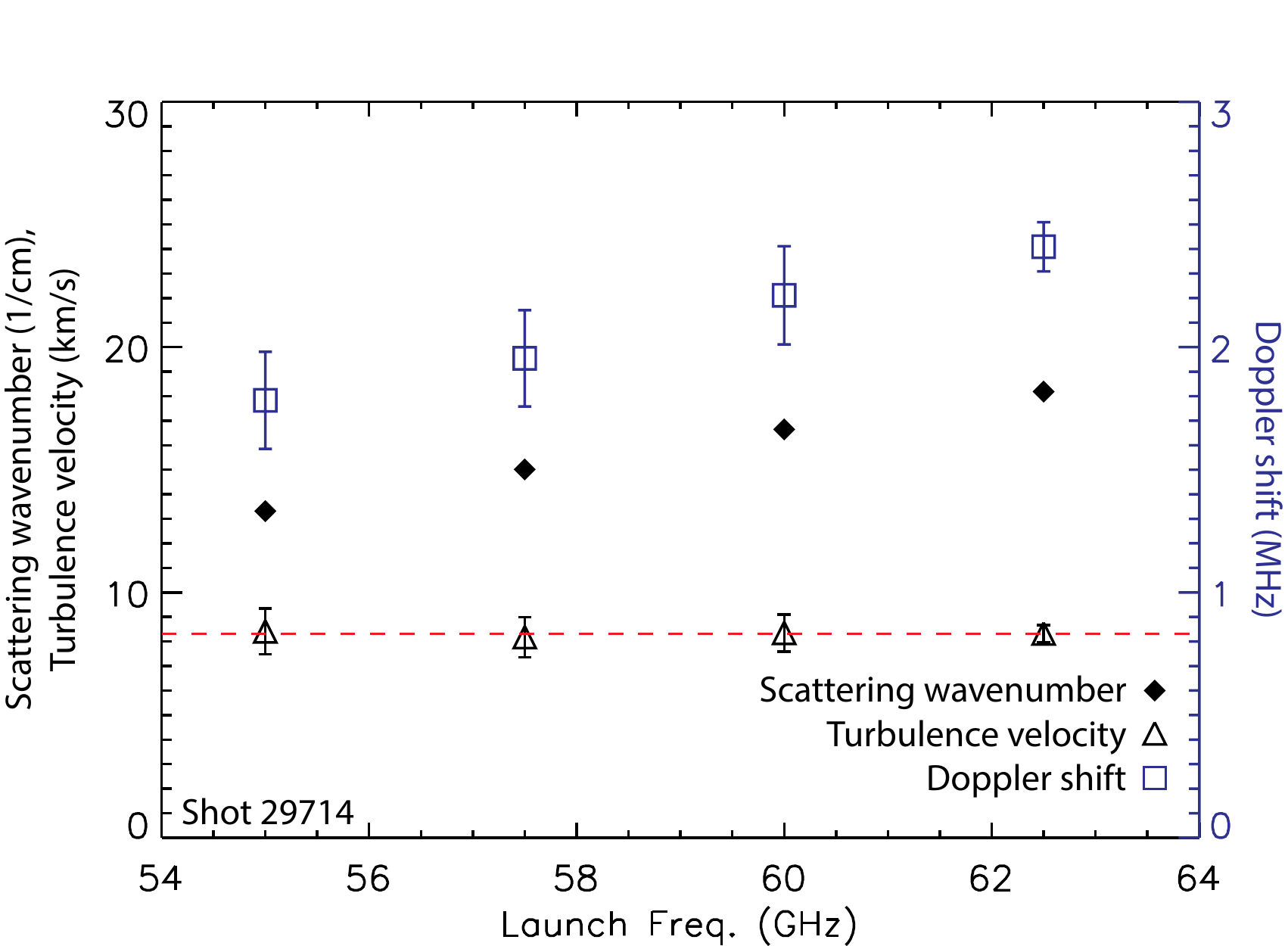}
\caption{\label{fig:loc1} Comparison of scattering wavenumber of the turbulence, turbulence velocity, and Doppler shift frequency for 4 channels with high-k trajectories localized together in radius, at 380 ms in shot 29714.  Dashed horizontal line is the mean value of the four turbulence velocity measurements.}
\end{figure}

Figure~\ref{fig:loc1} combines the measured Doppler shift from four channels of the V-band DBS system and ray tracing results for an Ohmic MAST discharge, shot 29714, at 380 ms.  Higher frequency channels, 67.5 GHz and above, for this shot had signals below noise levels.  Although there are in principle uncertainties introduced in the ray tracing, those should be systematic and not impact a relative comparison of the channels, so the error bars for the turbulence velocity are only propagated from the Doppler shifts.  This was a low current, $I_p$= 0.4 MA, and low density shot with line-averaged density at 380 ms of $\left< n_e \right> \approx 1.9 \times 10^{13}$ cm$^{-3}$.  At the low density the trajectory of the V-band channels, launched in O-mode polarization, were deflected little by refraction, so ray tracing results yielded that all four channels in Fig.~\ref{fig:loc1} had the lowest perpendicular index of refraction along their respective paths at about the same radius, $\sqrt{\psi} \approx 0.48-0.49$.  The horizontal dashed line is at the mean value of the four $v_{turb}$ measurements.  As one would expect if the measurements were localized, the turbulence velocity measured by the four independent channels is the same, within uncertainties.  For these cases, the relative perpendicular index of fraction of the beams at the scattering location was $k_i/k_0 \approx 0.6-0.7$.  The results also are consistent with the turbulence phase velocity being small compared to the $E \times B$ velocity (or scaling weakly with wavenumber, which would not be expected).  Given that the determination of the Doppler shift frequency and the ray tracing are completely independent calculations, the consistency of the determined values for the turbulence velocity is a strong confirmation that the measurement is local and, due to small deflection of the beam, the measurements are localized to nearly the same location.  Since it is the inferred velocity that is used for this comparison, it is further confirmation that the data for these high-k trajectories can be interpreted as usual for DBS data.

\subsection{High-k wavenumber spectrum of turbulence} \label{sec:highk_spec}

Having validated the design approach and interpretation, and demonstrated localization of the high-k measurements, we now use the high-k measurements to investigate one of the fundamental characteristics of a turbulent system, the wavenumber spectrum of the turbulence.  DBS measurements have previously been used to extract wavenumber spectra at several experiments~\cite{hennequin_fluctuation_2006,happel_scale-selective_2011,vermare_wavenumber_2011,schmitz_reduced_2012}.  The low magnetic field in MAST results in large gyroradii for particles, such that the high-k measurements are well above the ion scale, where electron temperature gradient (ETG) modes are thought to be important.  A series of studies using a high-k scattering diagnostic in NSTX have investigated ETG turbulence~\cite{mazzucato_short-scale_2008,smith_observations_2009,ren_density_2011,yuh_suppression_2011,ren_experimental_2012,ren_electron-scale_2013}, but those measurements were sensitive primarily to the radial wavenumber, $k_r$, with $k_r>k_{\bot}$, while DBS measurements are sensitive to the binormal wavenumber, $k_{\bot}$ with $k_r \approx 0$ cm$^{-1}$. The latter is more directly relevant to both linear stability of the mode and to transport, since correlations of radial velocity fluctuations (which are related to the binormal gradient of potential fluctuations) with density and temperature fluctuations result in particle and heat transport.

In principle, the dependence of scattering efficiency on wavenumber can impact the inferred wavenumber spectrum of density fluctuations.  This has been investigated with fullwave simulations and found to be a small effect, $\sim 10\%$ or less, in the regime measurements are presented below~\cite{silva_global_2004,blanco_study_2008,lechte_investigation_2009}.  Multiple small angle scattering events have also been predicted to result in a non-linear saturation effect for DBS measurements~\cite{gusakov_multiple_2005}.  The criterion given for the onset of the non-linear regime is $\gamma=(\delta n/n)^2 k_0^2 \ell_d L_r \ln(\ell_d /L_r)>1$, where $\delta n/n$ is the density fluctuation level normalized to the density at the cutoff, $k_0$ is the vacuum wavenumber, $\ell_d$ is the distance traveled in the plasma, and $L_r$ is the radial correlation length of the density fluctuations.  Measurements of the density fluctuation level in other MAST plasmas put an upper bound on $\delta n/n$ in the measurement region discussed below of about $0.3 \%$~\cite{field_comparison_2014}, with $L_r \approx 5$ cm and $\ell_d \approx 40$ cm, this yields $\gamma \approx 0.8$ at 70 GHz.  As discussed in~\onlinecite{field_comparison_2014}, the bound of $0.3 \%$ is due to diagnostic sensitivity limits for BES. The actual fluctuation level is probably lower, particularly since below we focus on times before strong beam heating, so the DBS measurements here are likely farther into the linear scattering regime.


In general, the scattering alignment effects discussed earlier in this paper can matter significantly for the measured amplitude of the backscattered radiation, as shown in Fig.~\ref{fig:flucts1}.  For comparing measurements, as is necessary to construct a wavenumber spectrum, this effect should be taken into account.  Rather than calculating and correcting for this effect, which is beyond the scope of this work, toroidal angle scans allow us to empirically determine the optimal scattering alignment by identifying the peak received power as a function of toroidal angle.  We employ this approach for the toroidal angle scan described above in Figs.~\ref{fig:traces2}-\ref{fig:scan2}.  This sequence of shots was chosen since the first $\sim 200$ ms of the shots were repeated, and the low density resulted in the V-band system, which was operating in O-mode polarization, having high-k trajectories such that all 8 channels were localized $\sqrt{\psi} \approx 0.35-0.40$.  This allowed a wavenumber spectrum to be constructed, since there was a large variation in frequency and therefore scattering wavenumber.  Data above system noise levels was acquired on the 6 lowest frequency channels during the repeated time window of the sequence of shots.  For the highest frequency channels with analyzable DBS data, the toroidal angle scan isolated scattering alignment to a single toroidal launch angle. For $k_{\bot} > 20$ cm$^{-1}$ if the toroidal launch angle was varied by 1 degree in either direction the received DBS signal was below system noise levels.  Since all frequencies had similar trajectories, this identified the best scattering alignment for all channels.

\begin{figure}[!htbp]
\includegraphics[width=8.5 cm]{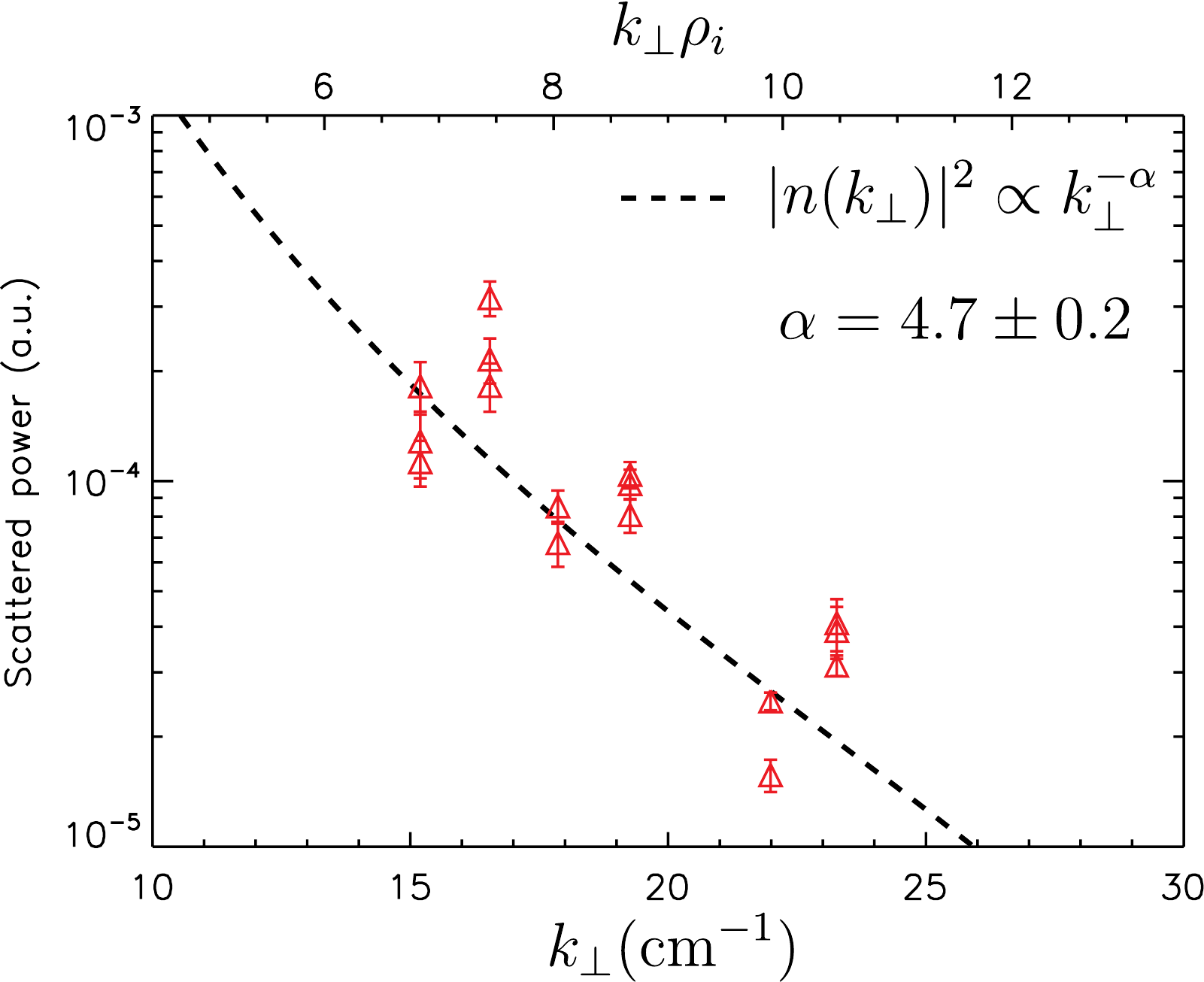}
\caption{\label{fig:highk_kspec} Measured dependence of density fluctuation power on scattering wavenumber at $\sqrt{\psi} \approx 0.35-0.40$.}
\end{figure}

Figure~\ref{fig:highk_kspec} shows the resulting wavenumber spectrum of density fluctuations inferred from the measurements.  Data are from three MAST shots, 29683, 29692, and 29693, which had the same toroidal and poloidal launch angles, averaged over 154-156 ms (just after NBI is added at 150 ms); there is about a factor of 2 variation between shots.  The general trend is clear, where there is more than an order of magnitude reduction in scattered power with about a $50 \%$ increase in scattering wavenumber.  Analyzing the Doppler shift for each channel in a similar way to Fig.~\ref{fig:loc1} reproduces the same result, confirming localization of the measurement.  Laboratory measurements were made of the relative response of each channel (which was accounted for), but \textit{in situ} calibration was not available; it is believed this is responsible for the apparent channel-to-channel variation.  The variation between channels might impact the inferred spectral index, but is not large enough to change the overall result.

Taken as a whole, the data set shows a strong reduction of density fluctuations with increasing wavenumber.  The cascade of energy in turbulent systems is usually governed by a power law.  Fitting the available data to a power law, $|n(k_{\bot})|^2 \propto k_{\bot}^{- \alpha}$, yields a spectral index $\alpha=4.7 \pm 0.2$.  Note that DBS is sensitive only to fluctuations with $k_r \approx 0$, which yields a different result than the spectrum integrated over all $k_r$~\cite{casati_turbulence_2009}.  These observations at $k_{\bot} \approx 15-23$ cm$^{-1}$ are at much smaller normalized scale than other reported DBS wavenumber spectra due to the low magnetic field in MAST, so there is not a clear comparison available from other experimental work.  Here, $k_{\bot} \rho_i \approx 7-11$ and $k_{\bot} \rho_e \approx 0.1-0.2$ (ion temperature from CXRS; electron temperature from Thomson scattering).  Comparison to theoretical expectations is discussed in the next section.

\section{Discussion and Conclusions \label{sec:conclusion}}

Doppler backscattering was successfully implemented for MAST.  Using considerations from scattering theory for design optimization, DBS was implemented using 2D steering.  We showed with data from the implementation that the dependence of measurements on toroidal alignment was consistent with expectations and provides measurements of wavenumber resolved density fluctuations and plasma flow.  Significantly, it was demonstrated that with toroidal steering to optimize alignment, DBS can measure high-k density fluctuations in spherical tokamak, $k_{\bot} \rho_i >10$, well into the electron scale.  We have also compared DBS flow measurements to measurements from charge exchange recombination spectroscopy and beam emission spectroscopy.  We found good agreement in most cases, except inside an internal transport barrier, consistent with a large poloidal rotation velocity associated with the ITB.  The inferred poloidal velocity was much larger than would be expected from neoclassical calculations that assume a small orbit width, motivating large orbit width effects to be included in future modeling and predictions.

We have shown how varying the toroidal launch angle impacts DBS measurements.  While this presents an issue in some cases for data interpretation, it in principle shows that toroidal steering for DBS can be used to provide information about the magnetic field configuration.  A different method for using microwave diagnostics to gain information about the magnetic field pitch angle through modifications to the beam profile has previously been discussed~\cite{gourdain_assessment_2008,gourdain_application_2008}.  It should in principle be possible to use data like that in Fig.~\ref{fig:flucts1} to constrain equilibrium reconstruction; that is, the peak scattered power provides a local constraint on the magnetic field pitch angle.  However, such a solution would also rely on the density profile and would require an iterative approach.  This might still be an attractive solution for future burning plasma devices, where microwave diagnostics are more robust to high neutron flux and power flux conditions than spectroscopic measurements like MSE.  

Finally, we have used the high-k measurement capability to investigate electron-scale turbulence.  The capability to steer the launched beam toroidally, to optimize the scattering alignment, was essential for probing high wavenumber density fluctuations.  In the core of an L-mode plasma, $\sqrt{\psi} \approx 0.35-0.40$, we found the power spectrum of density fluctuations reduced strongly with increased wavenumber, with $|n(k_{\bot})|^2 \propto k_{\bot}^{- \alpha}$ and $\alpha=4.7 \pm 0.2$, in the range $k_{\bot} \rho_i \approx 7-11$ ($\rho_e \approx 0.1-0.2$).  The measured DBS spectrum corresponds to $n(k_{\bot}, k_r \approx 0)$, whereas most theoretical investigations consider $n(k_{\bot})$ integrated over all $k_r$.  Nonetheless, specific predictions exist for the power spectrum of electric potential fluctuations (for electrostatic turbulence, density fluctuations should scale similarly) for ITG turbulence and transitions between regimes in the inertial range~\cite{schekochihin_gyrokinetic_2008,barnes_critically_2011}.  These predictions for the wavenumber spectrum describe the physics of the kinetic entropy cascade and should be independent of the details of the drive.  It is predicted that between the drive scale (typically $k_{\bot} \rho_i \sim 0.3-0.5$ for ITG turbulence) and $k_{\bot} \rho_i \sim 1$ that $E_\varphi (k_{\bot}) \sim k_{\bot} \rho_i |\varphi_k|^2 \propto k_{\bot}^{-7/3}$, and that between $k_{\bot} \rho_i \sim 1$ and the collisional dissipation scale that $E_\varphi (k_{\bot}) \propto k_{\bot}^{-10/3}$.  The collisional dissipation scale is predicted to be 
\begin{equation} \label{eqn:dis}
(k_{\bot} \rho_i)_c \approx q^{1/5} \left(\frac{R}{L_{T_i}} \right)^{4/5} \left( \frac{v_{th,i}}{\nu_{ii} R} \right)^{3/5},
\end{equation}
where $L_{T_i}^{-1}=-\partial \mathrm{ln} T_i / \partial r$ is the inverse ion temperature gradient scale length, $v_{th,i}=\sqrt{2 T_i/m_i}$ is the ion thermal speed, and $\nu_{ii}$ is the ion-ion collision frequency. Evaluating Eqn.~\ref{eqn:dis} for the parameters of the measurements in Sec.~\ref{sec:highk_spec} yields $(k_{\bot} \rho_i)_c \approx 35$, which is larger than the measurements, so if only ion-scale turbulence drive were present, and kinetic electron and electromagnetic effects are not important for the cascade, the $E_\varphi (k_{\bot}) \propto k_{\bot}^{-10/3}$ scaling would be expected to apply.  The prediction of $\varphi_k^2 \propto k_{\bot}^{-13/3}$ ($-13/3 \approx -4.333$) is only slightly weaker than the best fit to the data of $\alpha=4.7 \pm 0.2$.  Diagnostic effects such as scattering efficiency and variations between channels are of plausible size to contribute to the relatively small difference.  It is notable that the high-k wavenumber spectrum of density fluctuations being due to the tail of a turbulent cascade from large scales, rather than from turbulence driven directly at small scales, is also a consistent interpretation of measurements from NSTX~\cite{ren_electron-scale_2013}, although the measurements there were $k_{\bot} \rho_s \sim 2-4$ and $k_r \rho_s \sim 5-13$ (where $\rho_s$ is the ion sound gyroradius) while here $k_{\bot} \rho_i \sim 7-11$ and $k_r \approx 0$.

It has also been predicted that dissipation at all scales can occur through energy transfer to stable eigenmodes~\cite{hatch_role_2009}, and that the parameter $L_{T_i}/L_c$, where $L_c=v_{th,i}/\nu_{ii}$ characterizes the collisional mean free path, is important for determining the saturation regime~\cite{hatch_transition_2013}.  Ref.~\onlinecite{hatch_transition_2013} predicts dissipation at small scales to dominate for small $L_{T_i}/L_c$ (e.g. the predictions from ~\onlinecite{schekochihin_gyrokinetic_2008,barnes_critically_2011} would apply) and dissipation at large perpendicular scales to be important for large $L_{T_i}/L_c$, with a transition around $L_{T_i}/L_c \approx 10^{-3}$.  The experimental value of $L_{T_i}/L_c$ in this case is about $2 \times 10^{-3}$, which would be consistent with dissipation at large (\textit{i.e.} above the collisional dissipation scale) perpendicular scales contributing to a steeper spectral index.  With two possible small effects expected to contribute to making the measured spectral index larger (measuring only $k_r=0$ and scattering efficiency wavenumber dependence) and the small difference between  $\alpha=4.7 \pm 0.2$ and 13/3, we cannot conclude whether the predicted dissipation through energy transfer to damped modes makes a significant contribution.  For a better comparison to theory non-linear gyrokinetic simulations for the experimental conditions will be required.  Future parametric experimental studies are also motivated.  

\acknowledgments
Thanks to E. Gusakov, F. Parra, M. Barnes, and S. Newton for useful discussions.  This work was supported by the RCUK Energy Programme under grant EP/I501045, HAS, the European Union's Horizon 2020 programme, and the US Department of Energy under DE-FG02-99ER54527.  To obtain further information on the data and models underlying this paper please contact PublicationsManager@ccfe.ac.uk.  The views and opinions expressed herein do not necessarily reflect those of the European Commission.  JCH's work partly supported through the Culham Fusion Research Fellowship and EFDA Fusion Research Fellowship programmes.  The authors thank the CCFE Team at MAST and C. Wannberg and X. Nguyen at UCLA for the range of technical and engineering support that enabled the successful implementation, and thank PPPL for their support of the collaboration and equipment transfer.  Dr. Terry Rhodes is also gratefully for acknowledged for allowing the Inspect spectral analysis program to be ported for use with MAST data.


\bibliographystyle{aipnum4-1}

\end{document}